\documentclass{aa}  

\usepackage{graphicx}
\usepackage{txfonts}
\usepackage{lipsum}      
\usepackage{lscape}            
\usepackage{placeins}           
\usepackage[colorlinks=true,linkcolor=blue,filecolor=blue,citecolor=blue,urlcolor=blue]{hyperref}                                

\newcommand{\grs}{{GRS~1915+105}\xspace}
\newcommand{\NICER}{{\em NICER}\xspace}
\newcommand{\RXTE}{{\em RXTE}\xspace}

\begin{document}

   \title{Compelling evidence of a link between the lags of the quasi-periodic oscillations and the radio jet in the black-hole X-ray binary GRS 1915+105}
   \titlerunning{Compelling evidence of a link between the lags of QPOs and the radio jet in \grs}

   \author{Candela Bellavita\inst{1,2,3}
        \and Mariano Méndez\inst{3}
        \and Federico García\inst{1,2}
        \and Sara Motta\inst{4}
        \and Pengcheng Yang\inst{3}
        \and Wyncko Tonckens\inst{3}        \and Liang Zhang\inst{5}
        \and Rob Fender\inst{6}
        }

  \institute{Instituto Argentino de Radioastronom\'ia (CCT La Plata, CONICET; CICPBA; UNLP), C.C.5, 1894 Villa Elisa, Argentina
   \and Facultad de Ciencias Astron\'omicas y Geof\'isicas, Universidad Nacional de La Plata, 1900 La Plata, Argentina
         \and
              Kapteyn Astronomical Institute, University of Groningen, PO BOX 800, NL-9700 AV Groningen, the Netherlands
          \and
              Istituto Nazionale di Astrofisica, Osservatorio Astronomico di Brera, via E. Bianchi 46, 23807 Merate (LC), Italy
        \and
              Key Laboratory of Particle Astrophysics, Institute of High Energy Physics, Chinese Academy of Sciences, Beĳing 100049, China
          \and
          Astrophysics, Department of Physics, University of Oxford, Keble Road, Oxford OX1 3RH, UK
             }

   \date{Received September 30, 20XX}

  \abstract
  % context heading (optional)
  % {} leave it empty if necessary  
   {\grs is one of the most studied black-hole X-ray binaries, known for its extreme variability and rich phenomenology. Previous studies of this source with the Rossi X-ray Timing Explorer reported a transition of the phase-lags of type-C quasi-periodic oscillations (QPOs) from soft, where low-energy photons lag the high-energy ones, when the QPO frequency is higher than $\sim 2$~Hz, to hard when the QPO frequency decreases below $\sim2$~Hz. The hard-lags of the QPO coincide with episodes of strong radio emission. We analyse \NICER observations of \grs obtained between 2018 and 2020, during a period in which the source flux decreased steadily, and perform a detailed spectral-timing study of the detected type-C QPOs. We find a type-C QPO with frequencies in the range of  $\sim1.3-3.9$~Hz, which displays soft lags and, contrary to the \RXTE observations, shows no evidence of hard phase lags at frequencies below 2~Hz. Quasi-simultaneous AMI-LA radio observations show consistently low radio flux ($\lesssim5$~mJy) during this period. These results appear to show that the hard QPO lags in \grs are linked to the presence of strong radio activity, suggesting that the relativistic jet is responsible for the hard phase lags, supporting a scenario in which QPO phase lags trace changes in coronal geometry and accretion–ejection coupling in \grs.}

   \keywords{ X-ray: binaries -- X-ray: individual (GRS 1915 + 105)
               }

   \maketitle
   \nolinenumbers

\section{Introduction}

 Black-hole X-ray binaries (BHXBs) are binary systems consisting of a stellar-mass black hole accreting matter from a companion star \citep[for reviews, see][]{Remillard2006,Belloni2016}. Most BHXBs are transient sources, spending long periods in quiescence and occasionally undergoing outbursts of active accretion during which their X-ray luminosity increases by several orders of magnitude \citep{fender2009,dunn2010,uttley2015,Tetarenko2016}. These outbursts are accompanied by systematic changes in the spectral and timing properties of the source that define a sequence of accretion states, commonly referred to as low-hard state (LHS), hard-intermediate state (HIMS), soft-intermediate state (SIMS) and high-soft state (HSS) \citep{Homan2001,Belloni2005,Homan2005}. The X-ray spectra of BHXBs are usually described by a thermal component emitted by an optically thick, geometrically thin accretion disc \citep{ShakuraSunyaev1973}, and a hard component, produced by Comptonisation in a plasma of highly energetic electrons \citep{SunyaevTitarchuk1980}, called the corona. The spectral states are determined by changes of the physical properties and relative contribution of these components to the total flux. However, the geometry of the corona and its coupling to the disc and the jet remain key open problems.

BHXBs exhibit high X-ray variability over a broad range of timescales, from tens of milliseconds to years \citep{mendez1997,dunn2010,munozdarias2011,Motta2016}, providing a powerful probe of the accretion close to the black hole. Both the broadband noise and quasi-periodic oscillations (QPOs) encode information about the physical processes and geometry of the emitting regions \citep{vanderklis1994,vanderKlis2006,Fender2004,Ingram2009,Belloni2011,Kara2019,Mastroserio2019}. In particular, low-frequency QPOs (LFQPOs), observed in the range from a few millihertz up to tens of hertz \citep{Belloni2002,Casella2004,Remillard2006,Motta2012}, are a ubiquitous feature of BHXBs and show strong correlation with spectral state and luminosity.

LFQPOs are classified into three types, A, B, and C, based on their centroid frequency, quality factor, fractional rms amplitude, strength of the associated broadband noise, and phase-lag properties  \citep[][for reviews, see \citealt{Motta2016,Belloni2016}]{Wijnands1999, Remillard2002, Casella2004,Casella2005}. Type-C QPOs are the strongest LFQPOs with rms amplitudes that can reach up to $20\%$. They are typically detected in the LHS and HIMS \citep{Casella2004,Motta2011}, and their centroid frequency evolves smoothly with the position of the source in the hardness-intensity diagram (HID) in the range of $0.1-30$~Hz \citep{Casella2004,Belloni2005}, strongly correlated with the source spectral hardness \citep{Motta2011}. Type-B QPOs are associated with the SIMS \citep[][but see also \citealt{Jin2026}s, where the type-B QPO appears to be present in the HIMS]{Motta2011}, show rms amplitudes below $5\%$, are accompanied by a significant drop of the broadband variability of the source, and cluster around 6~Hz \citep{Casella2004,Casella2005}. Type-A QPOs are the least common and the weakest LFQPOs, appearing in the SIMS and HSS, with frequencies of $6-8$~Hz \citep{Belloni2014}. 

Phase lags between correlated variability in different energy bands add an additional diagnostic of the physical and geometrical properties of the inner accretion flow. The lags are said to be hard when the high-energy photons lag the low-energy ones, and soft when the opposite occurs. It has been proposed that hard and soft lags reflect either the Comptonisation delays with feedback between the corona and the disc \citep{Lee2001,reig2003, Giannios2004,Karpouzas2020, Garcia2022}, propagation of accretion-rate fluctuations \citep{arevalo2006,Ingram2013, Wilkinson2009}, or reverberation \citep{demarco2015, Kara2019}. For type-C QPOs, the energy dependence and sign of the lags provide direct constraints on the Comptonising region and its coupling to the disc and, potentially, to the jet \citep{eijnden2017,Mendez2022,Garcia2022}.

\grs is a very particular and complex BHXB that was first detected during outburst in August 1992 and has been active ever since \citep{castrotirado1992,castrotirado1994}. \grs was the first Galactic source in which relativistic superluminal radio ejections were observed \citep{Mirabel1994}. Jets are a fundamental outcome of accretion onto black holes, since they carry a significant fraction of the accretion power in collimated outflows that interact with the surrounding medium \citep{Fender2004}. Recent work has shown that the fastest jets tend to propagate along a fixed axis, while slower jets can vary in direction or precess \citep{Fender2025}. \grs remains an archetypal system for studying the coupling between accretion and relativistic-jet production.

Until July 2018, this source exhibited extraordinary variability and phenomenology that did not follow the canonical q-shaped hardness-intensity diagrams \citep{Belloni2016} observed in most transient BHXBs \citep{Homan2001,Fender2004, belloni2010}. Instead, until 2018, \grs cycled through three source-specific states, conventionally labelled A, B, and C, which only partially map onto the standard hard, intermediate, and soft states \citep{belloni2000}. State C is spectrally hard, shows strong aperiodic variability, and is commonly associated with steady radio jets \citep{rushton2010}, while states A and B are softer and show weaker variability \citep{belloni2000}. 

Type-C QPOs are a defining feature of state C in \grs and have been studied extensively \citep{Reig2000,Pahari2013, eijnden2016}. A remarkable result from observations with the Rossi X-ray Timing Explorer (\RXTE) is that the phase lags of these QPOs change sign as a function of QPO frequency. Several studies reported a transition from soft lags at high QPO frequencies to hard lags when the QPO frequency drops below $\sim$2 Hz \citep{Reig2000,Qu2010,eijnden2017}. \cite{Zhang2020}, analysing more than 600 \RXTE observations, show that the slope of the QPO phase-lag energy spectrum is negative (soft lags) above $\sim$2~Hz, consistent with zero around that frequency, and becomes positive (hard lags) at lower frequencies.

This behaviour was further investigated by \cite{Mendez2022}, who combined \RXTE timing data with nearly daily radio monitoring \citep{Pooley1997}. They demonstrated that the appearance of hard QPO lags below $\sim$2 Hz coincides with episodes of strong radio emission, directly linking the QPO lag behaviour to the presence of a jet. In that framework, soft lags are interpreted as arising from a compact, disc-coupled corona with strong radiative feedback, while hard lags signal a geometric transition in which the corona expands vertically and part of the coronal plasma is redirected into the jet. In this picture, the 2-Hz QPO frequency marks a sharp transition in coronal geometry and accretion–ejection coupling.

In 2018, after more than two decades of persistent activity, \grs entered a prolonged decay phase with a dramatic drop in both X-ray and radio flux \citep{Negoro2018, Motta2019}. During this phase, the source hardened spectrally and showed little or no radio emission \citep{motta2021,Zhou2025}. In May 2019, \grs exhibited flaring activity at different frequencies \citep{Motta2019, Koljonen2019, motta2021} that marked the beginning of an X-ray obscured state \citep{miller2020}, which is believed to be the result of partial-covering absorption produced by accreting and outflowing material. Observations with the Neutron Star Interior Composition Explorer \citep[\NICER;][]{Gendreau2016}, with its soft X-ray coverage and dense sampling, provide a unique opportunity to test whether the lag–frequency connection identified with \RXTE persists when the jet is weak or absent, and whether the hard lags observed with \RXTE are indeed connected to the jet.

In this work, we analyse X-ray and radio observations of \grs obtained between 2018 and 2020, during a period in which a systematic decrease of the X-ray flux was observed, following 26 years in which the source was persistently bright. We focus on the properties of type-C QPOs detected during this period, when the source shows hard spectra, decreasing QPO frequencies, and suppressed radio emission. By combining X-ray spectral–timing analysis of observations with \NICER and quasi-simultaneous radio observations with the Arcminute Microkelvin Imager Large Array (AMI-LA) at 15.5 GHz, we test whether the hard-lag regime reappears when the QPO frequency drops below $\sim$2~Hz. This allows us to directly probe the role of the jet and the coronal geometry in shaping the phase-lag properties of type-C QPOs in \grs. In Sec.~2 we describe the observations and data reduction. In Sec.~3 we present the results obtained from the analysis of the \NICER data and their comparison with the radio data and previous \RXTE studies. Finally, in Sec.~4 we discuss the implications of our findings.

\section{Observations and data analysis}
\label{sec:data}
\subsection{{\em NICER} data}
\subsubsection{Data processing}
We examined all the \NICER observations of \grs during the dimming phase from April 2018 to December 2019 (MJD $58200-58900$). This period encompasses the decay of the 26-year-long outburst \citep{Zhou2025}, as well as the transition in May 2019 (MJD 58617) into the unprecedented `X-ray obscured state', characterised by heavy local absorption and renewed radio flaring \citep{motta2021}. We used the \texttt{nicerl2} tool to process the \NICER data of \grs and produce clean event files. 

We used the GHATS\footnote{\url{https://github.com/ghats-timing/ghats}} package to extract power spectra (PS) of each observation\footnote{An observation corresponds to a unique obsID number in the \NICER archive} in the $0.3-12.0$~keV band, with a time resolution of 0.4~ms over segments of 65.536~s, which yields a lowest frequency and a frequency resolution of 0.015~Hz and a Nyquist frequency of 1250~Hz. GHATS generates the Leahy-normalised \citep{leahy1983} PS for each segment, which are then averaged to produce the PS of the observation. To correct for the Poisson level in the PS, we subtracted the average power in the frequency range $400-800$~Hz. To increase the signal-to-noise ratio (S/N) at intermediate and high frequencies, we performed a logarithmic rebinning, with the size of each bin increasing by a factor $10^{1/100}$ compared to the previous bin. We finally normalised the PS to fractional rms units \citep{belloni1990}, ignoring the background since its contribution to the total count rate is negligible. 

For the subsequent analysis, we excluded observations with fewer than 10 segments of $\sim 65$~seconds in the PS or with count rates below 100 counts/s. We also calculated the broadband fractional rms in the range $0.015-20$~Hz and discarded observations with rms values below $5\%$. Using these criteria, we discarded observations with low S/N or low variability, in which QPOs cannot be detected significantly. These selections reduced the initial sample of 204 observations to 80.
We then extracted light curves in the $0.3-12$~keV energy band with 1~second time resolution. We only retained observations or segments of observations exhibiting variability patterns similar to those from the $\chi$ class of \citet{belloni2000}, and excluded light curves showing structured or large-amplitude variability characteristic of other classes (e.g. $\lambda$, $\kappa$ or $\nu$), as in these cases the QPO frequency can change significantly within an observation.

For the remaining data, we used \texttt{XSPEC v.12.15.0f} to fit the PS in the $0.3-12$~keV energy band with multiple Lorentzians.  We only kept observations in which at least one narrow QPO feature was detected with a significance above $3\sigma$. Finally, we computed the dynamical power spectra in the $0.3-12$~ keV energy band with segments of 65.536~s using \texttt{gh\_dyn\_fits} and discarded observations or filtered segments where the QPO frequency changed significantly. 

We used the tool \texttt{nicerl3-spec} to extract energy spectra in the $0.3-12$~keV of the observations in which we identified a QPO. To estimate the background, we used the \texttt{SCORPEON} model provided by the \NICER team. Finally, we discarded observation 1103010185 due to strong particle contamination. After all selections, our final sample consisted of 71 observations, which are listed in Table~\ref{tab:tabla} in Appendix~\ref{sec:appendix}.

\subsubsection{Spectral analysis}
We fitted the energy spectra in the $0.3-12$~keV of each observation with \texttt{XSPEC} using the simple phenomenological model \texttt{tbabs~$\times$~(nthcomp+diskbb)}, ignoring data at energies above 10~keV because in several observations the background was high due to particle contamination. The component \texttt{diskbb} accounts for the thermal emission from an optically thick and geometrically thin accretion disc \citep{Mitsuda1984,Makishima1986}, and it has two parameters: the temperature at the inner disc radius, $kT_{\rm bb}$, and the normalisation of the component, $N_{\rm dbb}$, defined as the cosine of the inclination of the disc with respect to the line of sight times the square of the ratio between the inner radius and the distance to the source. The component \texttt{nthcomp} models the Comptonisation continuum produced by the corona \citep{Zdziarski1996, Zycki1999}. The parameters of this component are the power-law photon index, $\Gamma$, the electron temperature of the corona, $kT_{\rm e}$, the temperature of the seed photons, $kT_{\rm s}$, and the normalisation. In our fits, we linked the temperature of the seed photons, $kT_{\rm s}$, to the temperature at the inner radius of the disc, $kT_{\rm bb}$. We fixed $kT_{\rm e}$ to 100~keV, as it could not be constrained due to the limited energy coverage of \NICER. A similarly high electron temperature is reported by \citet{Zhou2025}, who fit \NICER and Insight–Hard X-ray Modulation Telescope data simultaneously and find that during the decay phase $kT_{\rm e}$ reaches the upper limit of their allowed range ($\sim$150 keV). The absorption of X-rays by the interstellar medium was modelled with the component \texttt{tbabs}, using the abundances and cross-section given by \citet{Wilms2000} and \citet{Verner1996}, respectively. To account for the use of a simplified continuum model that excludes broad emission lines, absorption edges or reflection, we added a $2\%$ systematic error to the statistical error in quadrature.

\subsubsection{Broadband PS and CS}
Using GHATS, we computed the PS in two energy bands, $2.0-5.0$~keV and $5.0-12.0$~keV, and the cross spectrum (CS) between the same bands. We followed the same procedure described above for the computation of PS in the full energy band, adopting the same time resolution, segment length, rebinning factor, and normalisation. We excluded the data from $0.3-2.0$~keV because the source is highly absorbed in this energy range \citep{neilsen2020,Nathan2022}. Using GHATS we also computed the phase-lag frequency spectra and the coherence function. We took the energy band $2.0-5.0$~keV as the reference band when computing the CS and phase-lag frequency spectrum. 
Given that during these observations the phase lags of \grs are close to zero over a broad range of frequencies, we rotated all cross vectors by $45^\circ$. As explained in \citet{Mendez2024}, this rotation makes the fit of the CS more stable without changing the best-fitting parameters.

We used \texttt{XSPEC} to fit simultaneously the PS in the $2.0-5.0$~keV and $5.0-12.0$~keV bands and the real and imaginary parts of the CS in the same energy bands, considering the $0.01-50$~Hz frequency range. Following the technique in \citet{Mendez2024} assuming constant phase lags with Fourier frequency, we started by fitting a single Lorentzian function and we added a new Lorentzian at a time until there were no systematic trends in the residuals and the reduced $\chi^2$ was about unity. During the fits, we linked the centroid frequencies and FWHM of each Lorentzian in the two PS and in the CS but left the normalisations of the PS free to vary independently. For each Lorentzian, its normalisation in the CS was tied to the square root of the product between the two PS normalisations \citep[see][]{Mendez2024}. Each Lorentzian component included was at least $3\sigma$ significant in the PS and CS.

\subsubsection{Energy-resolved PS and CS}

To explore the fractional rms and phase-lag spectra of QPOs, we extracted the PS in seven subject energy bands: $0.3-2.0$~keV, $2.0-3.0$~keV, $3.0-4.0$~keV, $4.0-5.0$~keV, $5.0-6.0$~keV, $6.0-8.0$~keV, $8.0-12.0$~keV. We also generated the CS of each band with respect to the total band $0.3-12$~keV, which was therefore the reference band for the lags. To correct for the partial correlation of the photons that are simultaneously in the subject and the reference bands, we subtracted the average of the real part of the CS calculated in a frequency range where the source does not contribute to the variability \citep{Belloni2024,Mendez2024}. Using the best-fitting model obtained from the joint fit of the PS and CS for each observation, and fixing the centroid frequencies and FWHMs of every Lorentzian to the values obtained in those fits, we fitted the PS in each subject band simultaneously with the PS in the reference band and the real and imaginary parts of the CS between those two bands. We then constructed the rms amplitude and phase-lag spectra of the QPO in each narrow energy band.

\subsection{{\em RXTE} archival data}
\label{sec:RXTE}
For the purpose of this work, we do not reanalyse the \RXTE data. Instead, we use the published results to compare the frequency dependence of the QPO phase-lag spectra, the fractional rms behaviour, and their relation to the radio emission, as well as the spectral evolution through the photon index $\Gamma$.

We compare our results with those of \citet{Zhang2020}. They used observations of \grs obtained with the Proportional Counter Array (PCA) on board the \RXTE \citep{Zhang1993}. These observations span from 1996 to 2012 and provide a comprehensive reference sample of detected type-C QPOs. \citet{Zhang2020} computed the power spectra in the full PCA energy band, fitted them using a multi-Lorentzian model and measured the phase lags over the width of the QPO between the $2.0$--$5.7$~keV and $5.7$--$15.0$~keV energy bands. Finally, they calculated energy-dependent rms and phase-lag spectra for the QPO using the energy bands $4-6$~keV, $6-8$~keV, $8-11$~keV, $11-15$~keV, $15-21$~keV, and $21-44$~keV, taking the softest band $2-4$~keV as reference. \citet{Zhang2020} used the traditional method to calculate the QPO phase lags, which consists of averaging the CS over the frequency interval spanning the QPO width. This approach can be biased by the contribution of overlapping broadband variability components \citep{Mendez2024}. In this work, we instead used the multi-Lorentzian decomposition of \citet{Mendez2024}, which explicitly separates the QPO from the other variability components and obtains the phase lag of the QPO directly from the parameters of the fit. 
Because the QPO lag is estimated simultaneously with the parameters of all the other variability components, its uncertainty naturally reflects the uncertainties and covariances of the full model. Consequently, the resulting error is larger than that obtained with the traditional method, but it more faithfully represents the true uncertainty of the measurement. However, for the \RXTE observations analysed by \citet{Zhang2020}, the QPO strongly dominates the variability around its centroid frequency \citep{eijnden2016}, and both methods yield consistent phase-lag measurements, allowing the direct comparison of the results.

\citet{Mendez2022} performed spectral fits using the phenomenological model \texttt{vphabs~$\times$~(diskbb+gauss+nthcomp)}, from which they derived the photon index of the Comptonisation component, $\Gamma$, for each observation.

\subsection{AMI-LA data and Ryle data}
\label{sec:ami-la}
We complemented our analysis with radio data from the AMI-LA, obtained at a central frequency of 15.5 GHz with a 5-GHz bandwidth, divided into eight channels for imaging. The flux density scale was calibrated using 3C~286, while J1922+1530 was used as the phase calibrator. (See \citealt{motta2021} for more details about the reduction of the data.) The radio data were binned according to the source brightness. When the flux density was below $10~\mathrm{mJy~beam^{-1}}$, we considered the average flux density measured over each observing epoch, with durations ranging from 1 to 7~h. For brighter epochs, the observations were divided into shorter segments, down to 6~min, depending on the source flux. Quasi-simultaneous radio and X-ray observations were defined as those taken within 0.5~MJD with \NICER and AMI-LA.

To study the connection between X-ray timing properties and jet emission, \citet{Mendez2022} 
combined the \RXTE data with radio observations at 15~GHz obtained with the Ryle Telescope \citep{Pooley1997}. X-ray and radio observations were associated based on their observation dates, resulting in a sample of 410 observations with simultaneous X-ray timing and radio flux measurements. (See \citealt{Pooley1997} and \citealt{motta2021} for further details on the instrument and data reduction).

\section{Results}
\label{sec:results}

\begin{figure}
        \includegraphics[width=\columnwidth]{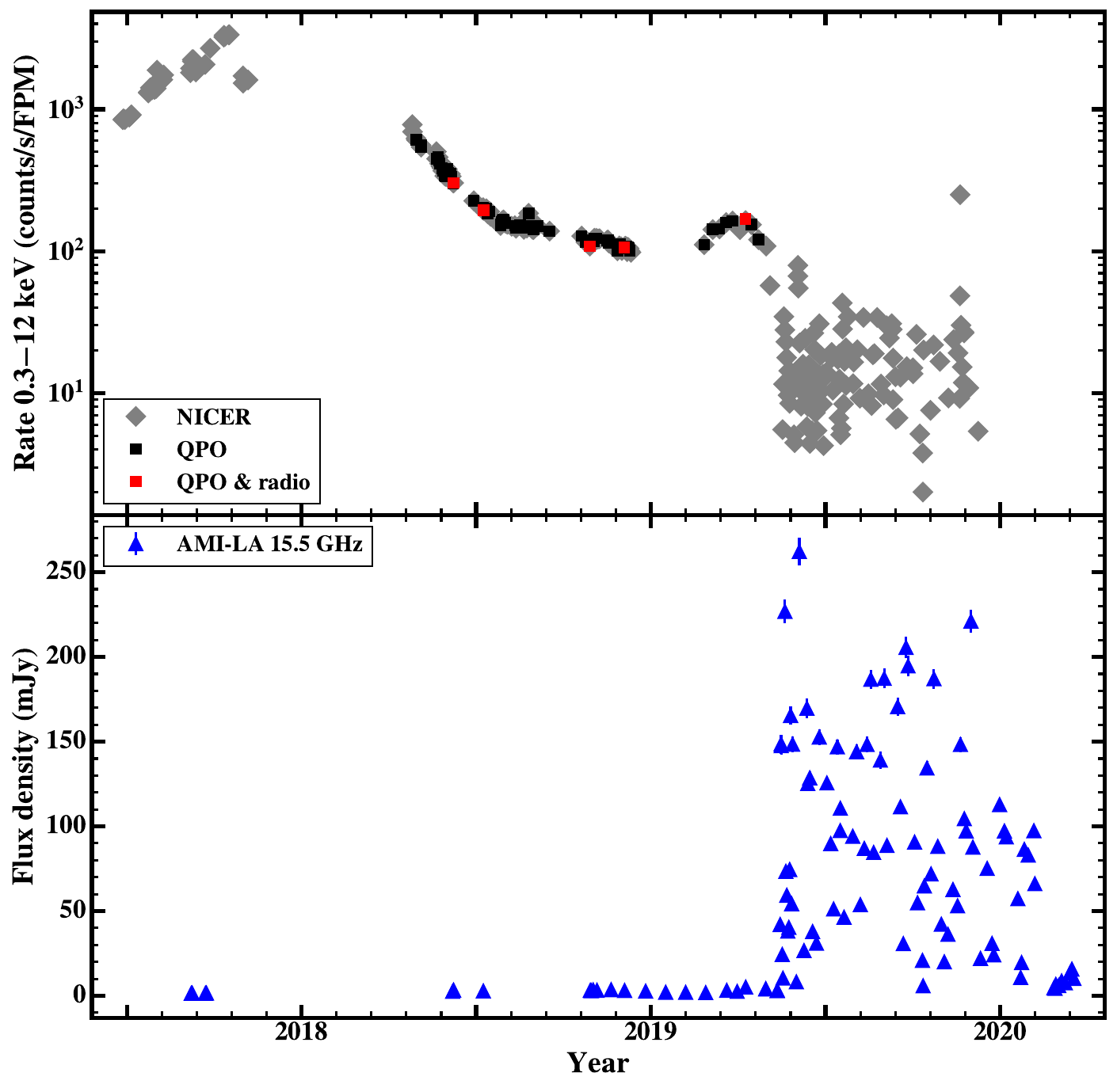}
        \caption{Top panel: \NICER X-ray lightcurve of \grs from June 2017 to December 2020. The intensity is the count rate per detector in the $0.3-12$~keV band. Each grey diamond corresponds to a \NICER observation, the black squares are the \NICER observations in which we detect a QPO, while the red squares are the observations with QPOs and a simultaneous radio measurement within 0.5 days. Bottom panel: AMI-LA radio 15.5 GHz light curve covering the same time period.}
        \label{fig:LC}
    \end{figure}

The top panel of Fig.~\ref{fig:LC} shows the \NICER light curve of \grs (grey diamonds) in the $0.3-12$~keV band from June 2017 to December 2019. The black squares indicate observations in which we detect a QPO. In the bottom panel of Fig.~\ref{fig:LC} we plot the 15.5~GHz AMI-LA data of the source in the same time interval. The \NICER data with quasi-simultaneous radio observations (taken within 0.5 days of the X-ray data) are highlighted with red squares in the top panel.

\begin{figure*}
        \includegraphics[width=\textwidth]{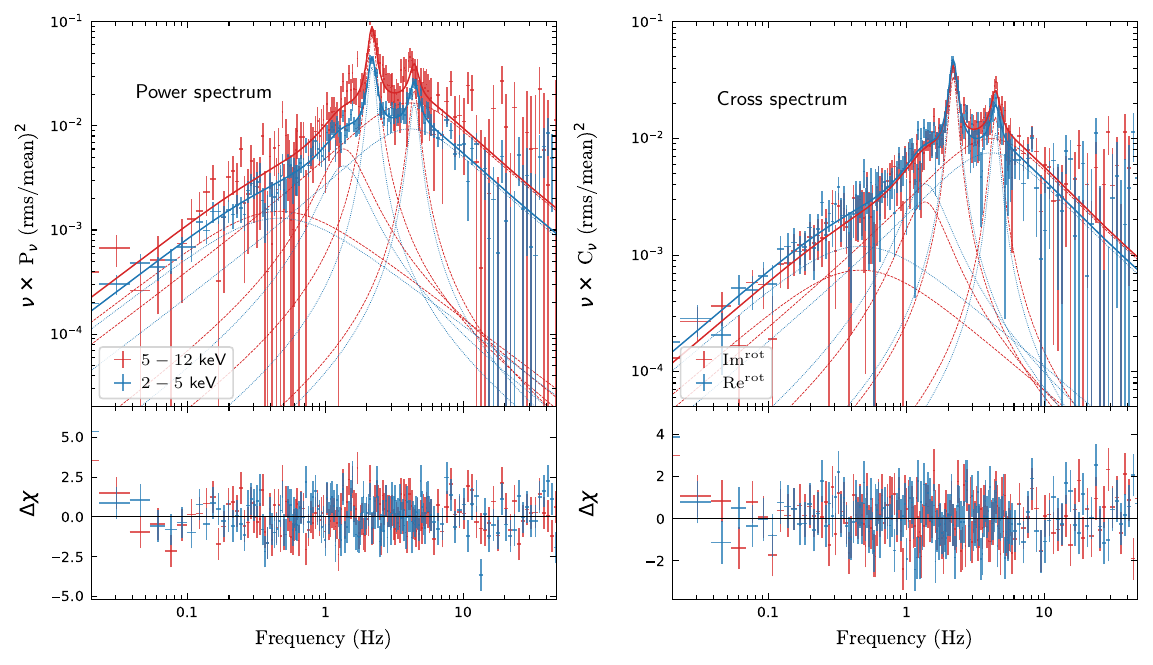}
    \caption{Left panel: Power spectra of \NICER observation 1103010158 of \grs in two energy bands. The $2-5$~keV data are shown in blue while the $5-12$~keV one is in red, both with the best-fitting model (solid line) consisting of 5 Lorentzian functions (dotted lines). Right panel: Real and imaginary parts of the cross spectrum between the same two energy bands rotated by $45^{\circ}$. We plot $\rm{Re}~cos(\pi/4) -\rm{Im}~sin(\pi/4)$ in blue and $\rm{Re}~sin(\pi/4) +\rm{Im}~cos(\pi/4)$ in red, with the best-fitting model assuming constant phase lags. In both panels, residuals with respect to the model, defined as $\Delta \chi = {\rm (data-model)/error}$, are also plotted. For visual purposes only, the data above 6~Hz have been rebinned by a factor of 3 in the plot.}
        \label{fig:PSCS}
    \end{figure*}

As an example, in Fig.\ref{fig:PSCS} we show the best-fitting model for the observation 1103010158. In the top panels we present the best-fit to the two PS (left) and the real and imaginary parts of the CS (right), with their corresponding residuals in the bottom panels. The best-fitting model consists of five Lorentzians and yields $\chi^2 =811.09$ for 819 d.o.f. In particular, we see a narrow component at $\sim 2.2$~Hz that we identify as the type-C QPO, a possible subharmonic at $\sim 1.3$~Hz and a second harmonic at $\sim 4.4$~Hz.

\begin{figure*}
   \sidecaption
   \centering
   \begin{minipage}{12cm}
   \includegraphics[width=12cm]{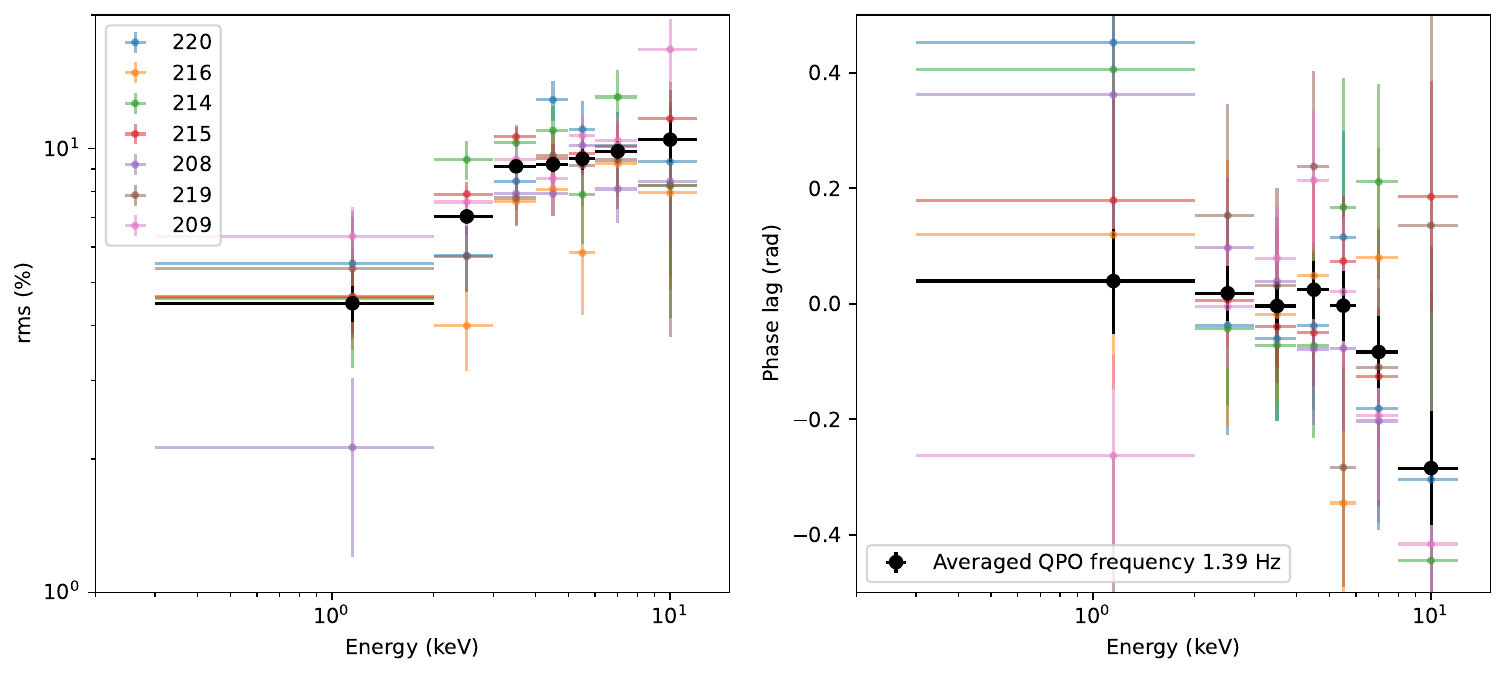}
     \vspace{0.5cm}
   \includegraphics[width=12cm]{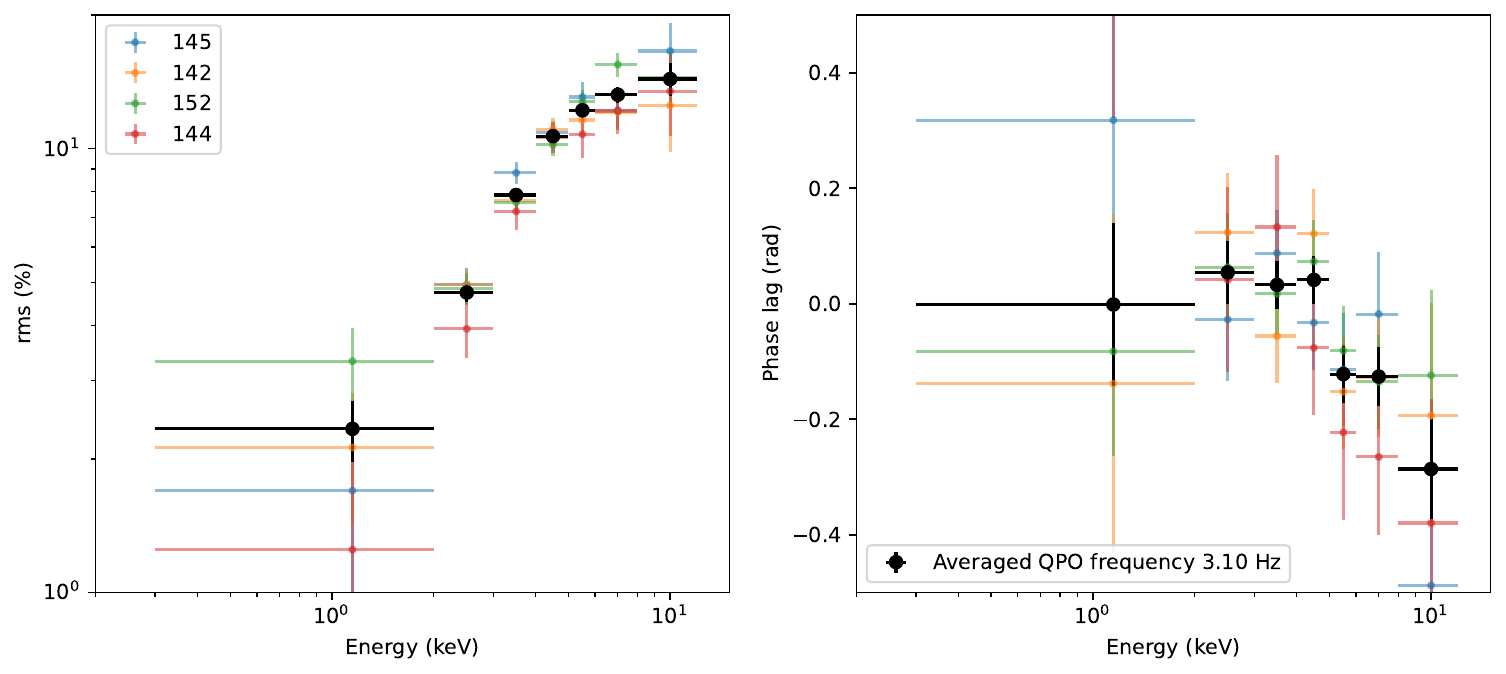}
   \end{minipage}
      \caption{Fractional rms (left panels) and phase-lag (right panels) spectra of the  QPO in \grs. The top and bottom panels correspond to different sets of observations. Filled black circles depict the weighted-averaged of the individual observations, which are marked with different colours. In the legends, the numbers indicate the last three digits of the observation IDs 1103010XXX. Horizontal error bars represent the energy range of each point. }
         \label{fig:rmsylagvsE}
   \end{figure*}

In Fig.~\ref{fig:rmsylagvsE} we show the rms (left) and phase-lags (right) spectra for several observations, depicted with different colours. The top panels present observations 1103010214, 1103010215, 1103010216, 1103010208, 1103010219, 1103010209 and 1103010220, corresponding to QPO frequencies in the range $1.36-1.41$~Hz. The bottom panels present observations 1103010145, 1103010142, 1103010152, and 1103010144, with QPO frequencies between $3.04-3.20$~Hz. The rms spectra across observations exhibit an increasing trend with energy while the lag spectra appear to decrease with energy. Given the large error bars of the individual phase-lag spectra, to investigate the overall trends, we computed weighted averages of both the rms and phase-lag energy spectra for observations with similar QPO frequencies. Within each frequency group, the individual spectra are consistent with each other within their uncertainties. The corresponding weighted-averaged QPO frequencies are 1.39~Hz and 3.10~Hz for the upper and lower panels, respectively. In both right panels, the averaged phase-lag spectra are soft, with QPO lags close to zero at low energies and decreasing down to~$\sim-0.3$~rad towards higher energies.

\begin{figure}
   \centering
   \includegraphics[width=\columnwidth]{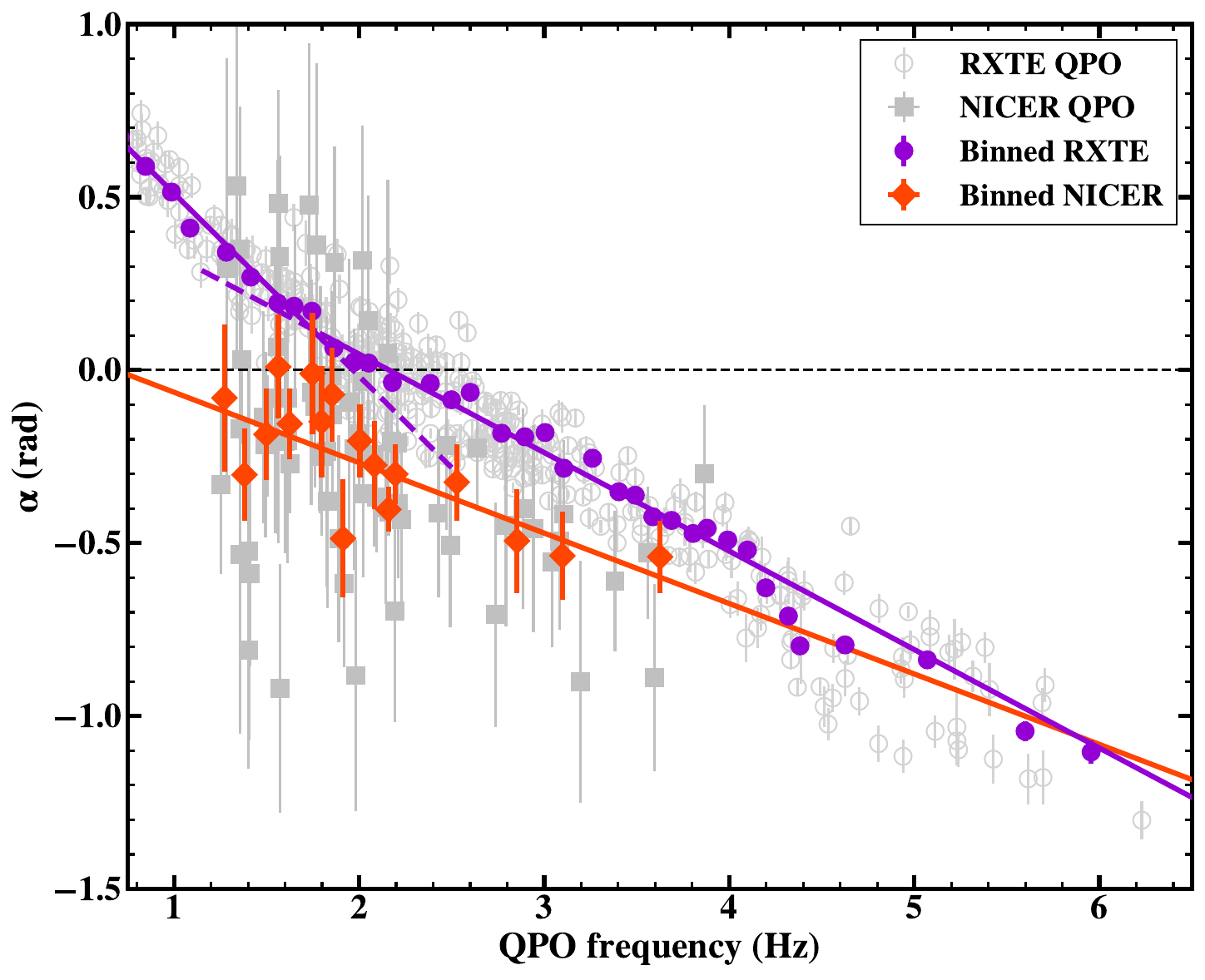}
      \caption{Slope of the phase-lag spectra of the type-C QPO in \grs as a function of the QPO  frequency. Empty grey circles and grey squares show individual \RXTE and \NICER data, respectively. The purple filled circles (\RXTE) and red diamonds (\NICER) show the data rebinned in frequency to reduce the error bars. The purple solid line shows the best-fitting broken line to the binned \RXTE data, with the purple dashed lines indicating its extrapolation for comparison. The red solid line shows the best fit to the \NICER binned data.}
         \label{fig:Slopevsnu}
   \end{figure}

The broadband lags are strongly dependent on the energy bands adopted. This is particularly relevant when comparing the lags of the QPOs in the \NICER data with those reported by \citet{Zhang2020} using \RXTE data, as we used the $2-5$~keV and $5-12$~keV bands, while \citet{Zhang2020} used the $2-5.7$~keV and $5.7-15$~keV bands. Therefore, following \citet{Zhang2020}, we fitted the phase-lag energy spectra of each QPO with the function $\Delta\phi(\rm{E})=\alpha log(\rm{E/E_{ref}})$. The slope of the lag-energy spectrum, $\alpha$, provides a band-independent characterisation of the energy dependence of the lags, and thus enables a direct comparison between the \NICER and \RXTE results. In Fig.~\ref{fig:Slopevsnu} we plot the $\alpha$ values with $1\sigma$ error bars against the QPO frequency. Empty grey circles and grey squares correspond to individual \RXTE and \NICER measurements, respectively. In the case of \RXTE, we fitted the logarithmic dependence of the phase-lag spectra with the energy using the $4-6$~keV, $6-8$~keV, $8-11$~keV, $11-15$~keV, $15-21$~keV, and $21-44$~keV energy bands, while for the \NICER data we used $2-3$~keV, $3-4$~keV, $4-5$~keV, $5-6$~keV, $6-8$~keV, $8-12$~keV energy bands. To test the effect of the lag measurement method, we calculated the phase-lag spectra using the traditional method adopted by \citet{Zhang2020}, i.e. by averaging the CS over the frequency interval spanning the QPO width, and fitted the resulting lag-energy spectra with the same logarithmic function. The slopes obtained with this approach are consistent, within the uncertainties, with those obtained from the multi-Lorentzian decomposition. To assess the impact of the softer \NICER bandpass on the measured slopes, we repeated the fit twice, first excluding the $2-3$~keV bin and then excluding also the $3-4$~keV bin, i.e. using only bins above 3~keV and above 4~keV, respectively. The resulting slopes are consistent, within $1\sigma$, with those obtained using the full $2-12$~keV band, although the uncertainties increase as the lower-energy cutoff is raised, due to the reduced statistics. We therefore consider the full $2-12$~keV band throughout the rest of this work and, since the uncertainties in the \NICER data are relatively large, we also plot the weighted-average slopes with red diamonds.
This figure shows that, as already noticed by \citet{Zhang2020}, in the \RXTE data the slope of the lag spectrum decreases as the QPO frequency increases, changing sign from positive to negative at around 2~Hz. \citet{Zhang2020} find that a broken line fits the correlation of the slope with the QPO frequency better than a straight line, so we independently binned the \RXTE measurements in QPO frequency (purple filled circles in Fig.~\ref{fig:Slopevsnu}) and fitted them with a broken line (purple solid lines, extended with dashed lines for comparison), obtaining a break at $1.71 \pm 0.05$~Hz and slopes of $-0.53 \pm 0.03~{\rm rad~Hz^{-1}}$ and $-0.28 \pm 0.01~{\rm rad~Hz^{-1}}$ below and above the break, respectively. In contrast, the binned \NICER data show a consistently negative slope over the entire frequency range, also below 2~Hz. We fitted the correlation of the \NICER binned slopes with frequency with a linear function and we obtained a slope of $-0.21 \pm 0.02~{\rm rad~Hz^{-1}}$ from the best fit.

\begin{figure*}
   \centering
   \includegraphics[width=\textwidth]{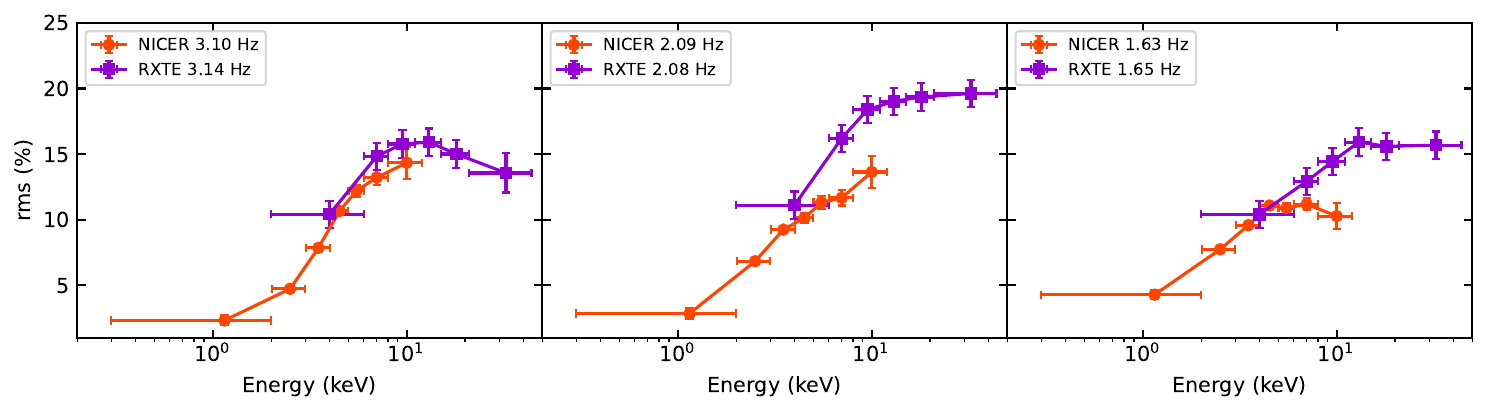}
      \caption{Comparison of the QPO fractional rms spectra of \grs measured with \RXTE and \NICER at three representative QPO frequencies. Purple circles show the rms spectra measured with \RXTE, while red squares show the \NICER rms spectra averaged over observations within the same frequency bins used in Fig.~\ref{fig:rmsvsslope}. From left to right, the panels correspond to QPO frequencies of approximately 3.1, 2.1, and 1.6~Hz.}
         \label{fig:rmsspectraRXTEvsNICER}
   \end{figure*}

The fractional rms measured over a broad energy range depends strongly on the adopted energy range and on the energy dependence of the instrument’s effective area. The effective areas of the two instruments have markedly different energy dependences over the considered energy ranges. As a result, photons at different energies contribute with different weights to the measured broad-energy-band variability, which may affect the measured fractional rms amplitudes. Before comparing the \RXTE and \NICER broadband rms amplitudes, we therefore estimated the effect of these differences by comparing the QPO fractional rms spectra of representative \RXTE observations with those of the weighted-average \NICER observations at similar QPO frequencies (Fig.~\ref{fig:rmsspectraRXTEvsNICER}). For QPO frequencies around 3~Hz, the RXTE and NICER rms spectra are consistent within the uncertainties over the common energy range (left panel of Fig.~\ref{fig:rmsspectraRXTEvsNICER}). We therefore fitted the RXTE and NICER rms spectra simultaneously with the same spectral shape, allowing only for a relative normalisation between the two instruments. Integrating the best-fitting models over the corresponding energy ranges yields an empirical scaling factor of $\sim1.6$ between the RXTE ($2-15$~keV) and NICER ($2-12$~keV) broadband rms amplitudes. This is only a crude empirical estimate, since the broadband rms depends not only on the instrumental response and energy range, but also on the source energy spectrum. Moreover, the scaling assumes that the agreement between the RXTE and NICER rms spectra extends into the $12-15$~keV energy range, where NICER has no coverage. At lower QPO frequencies, however, the NICER rms spectra remain systematically below those measured with RXTE, even over the common energy range, as can be seen in the middle and right panels of Fig.~\ref{fig:rmsspectraRXTEvsNICER}. Repeating the same procedure yields progressively larger normalisation factors as the QPO frequency decreases, suggesting that the difference between the RXTE and NICER broadband rms amplitudes cannot be explained solely by the different instrumental responses and energy bandpasses. This trend is nevertheless only marginally significant,  given the uncertainties and limited number of observations available for comparison.

\begin{figure}
   \centering
   \includegraphics[width=\columnwidth]{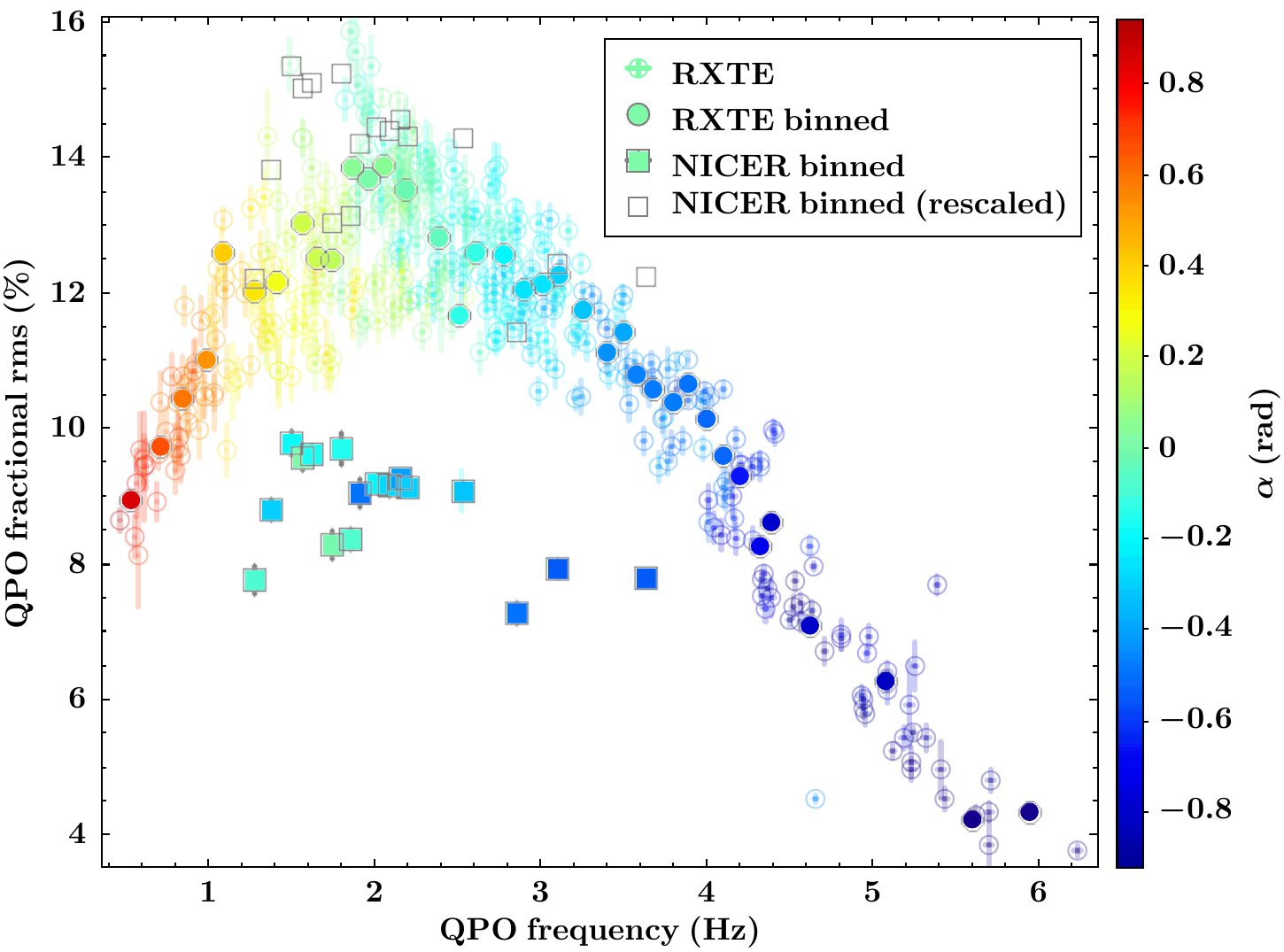}
   \caption{Fractional rms amplitude of the type-C QPO in \grs as a function of QPO frequency. Empty circles show the rms amplitudes measured from individual \RXTE observations in the $2-15$~keV energy band. Filled circles and filled squares show the weighted-average rms amplitudes of the \RXTE ($2-15$~keV) and \NICER ($2-12$~keV) observations, respectively, using the same frequency bins as in Fig.~\ref{fig:Slopevsnu}. The colour scale indicates the slope of the phase-lag spectrum, $\alpha$. Grey empty squares show the \NICER weighted-average rms amplitudes after applying the empirical scaling factor of $\sim1.6$ derived from the comparison of the \RXTE and \NICER rms spectra shown in Fig.~\ref{fig:rmsspectraRXTEvsNICER}.}
         \label{fig:rmsvsslope}
   \end{figure}

In Fig.~\ref{fig:rmsvsslope}, we present the QPO fractional rms amplitude as a function of QPO frequency, calculated in the $2-15$~keV band for \RXTE and in the $2-12$~keV band for \NICER. The empty circles show the individual \RXTE measurements, while the filled circles and filled squares show the weighted-average \RXTE and \NICER rms amplitudes, respectively, using the same QPO frequency bins adopted for the lag-energy slope analysis in Fig.~\ref{fig:Slopevsnu}. In Fig.~\ref{fig:rmsvsslope}, the colour scale represents the slope of the phase-lag spectra, $\alpha$, which we use as a proxy for the broadband lags, given that it captures both the sign and energy dependence of the lags in a manner that is less sensitive to the specific choice of energy bands. As we mentioned before, a direct quantitative comparison between the \RXTE and \NICER rms amplitudes should be treated with caution. As shown by \citet{Zhang2020}, the QPO fractional rms of the RXTE data first increases as the frequency of the QPO increases and the slope of the phase-lag spectrum decreases,  reaching a maximum fractional rms at around $\nu_{\rm QPO}\approx 2$~Hz and $\rm{slope}\approx 0$~rad. As the QPO frequency increases further, the QPO fractional rms decreases and the slope of the QPO phase-lag spectrum becomes negative. As discussed above, the systematically lower \NICER rms amplitudes can be partly explained by the different energy coverage and instrumental responses. Applying the empirical scaling factor of $\sim1.6$, derived from the comparison of the energy-dependent rms spectra between the two instruments, reduces the difference between the \RXTE and \NICER measurements (grey empty squares in Fig.~\ref{fig:rmsvsslope}.)

\begin{figure}
   \centering
   \includegraphics[width=\hsize]{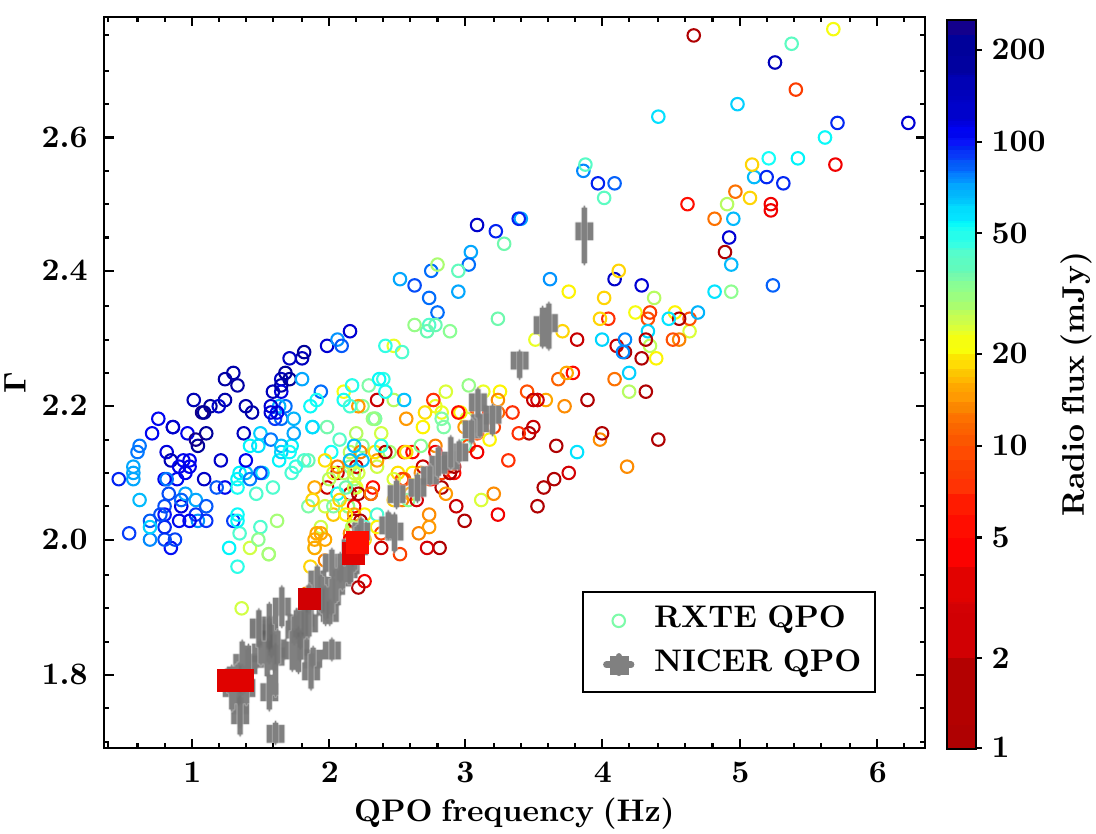}
      \caption{Photon index of the Comptonised component in \grs as a function of frequency of the type-C QPO. Open circles are the \RXTE data from \citet{Mendez2022} and \citet{Garcia2022}, while filled squares are the \NICER data with QPOs. The colours represent the 15.5~GHz radio flux in mJy. The \NICER observations without a simultaneous (within 0.5~days) radio observation are shown in grey.}
         \label{fig:Gammavsnu}
   \end{figure}

As a final step, we investigate the spectral evolution of the source during the epochs in which a QPO is detected. Given that hardness ratios are strongly instrument-dependent, we cannot compare the \RXTE and \NICER data in a plot of hardness ratio vs. QPO frequency, like the ones presented in \citet{Mendez2022}. Since the hardness ratio is strongly correlated with the power-law index of the Comptonisation component, $\Gamma$, \citep{Garcia2022}, we adopted  $\Gamma$ as a tracer of the spectral state, enabling a direct comparison between the \NICER and \RXTE observations. As we described in Sec.~\ref{sec:data}, for all the observations in which we detect a QPO, we fitted the energy spectra in the $0.3-10$~keV using a simple phenomenological model of \texttt{tbabs~$\times$~(nthcomp+diskbb)}, with the purpose of capturing the overall shape of the continuum rather than providing a detailed physical description \citep[see][for a similar approach]{Konig2024,Fogantini2025}. In Fig.~\ref{fig:Gammavsnu} we plot the photon index $\Gamma$ against the frequency of the QPO for the \NICER observations (filled squares). The photon index increases from 1.7 to 2.4 when the QPO frequency increases from $\sim 1.2$~Hz to 3.9~Hz. For comparison, we also show the \RXTE measurements (open circles) from \citet{Mendez2022}, colour-coded according to the radio flux density measured with the Ryle Telescope at 15.5~GHz. In the \RXTE data, we identify two branches, one for high radio fluxes, above $\sim60$~mJy, and one for low radio fluxes, below $\sim 20$~mJy. We notice that the \NICER data aligns with the low radio flux branch. The \NICER observations with an AMI-LA radio observation taken within 0.5~days, marked by red squares, correspond to radio flux densities below $\sim5$~mJy. Although the grey points do not have simultaneous radio measurements, Fig.~\ref{fig:LC} shows that \grs was radio quiet during the whole period of our \NICER observations, indicating that they most likely also correspond to low radio fluxes.

\section{Discussion}
\label{sec:discussion}
We investigated the properties of type-C QPOs in \grs observed with \NICER between April 2018 and December 2019 during the period in which the X-ray flux of the source decreased steadily. We detect type-C QPOs with centroid frequencies between $\sim1.3$ and $3.9$ Hz whose phase-lag spectra are always consistent with being soft, but have large uncertainties. By combining observations in bins of QPO frequency to increase the signal-to-noise ratio, we find a significant anticorrelation between the lags and the QPO frequency, with the lags remaining systematically soft across the entire frequency range. This behaviour is in contrast with earlier \RXTE results for \grs, which, over the $\sim$16-year monitoring of the source, consistently showed a transition to hard lags whenever the QPO frequency decreased below $\sim2$~Hz. By combining our X-ray timing results with quasi-simultaneous AMI-LA radio observations, we find that the soft-lag regime observed with \NICER coincides with periods of weak radio flux, whereas the hard-lag regimes seen in \RXTE data are associated with strong radio activity. This comparison suggests that the hard lags of the QPO are closely linked to the radio jet.

\cite{Zhang2020} fitted the phase-lag energy spectrum of the type-C QPO in the \RXTE data using a log-linear function, characterised by a slope, $\alpha$, which switches from positive to negative as the QPO frequency increases, with the sign change happening at $\sim2$~Hz. They further show that $\alpha$ follows a broken-line relation with the break also at $\sim 2$~Hz and with slopes of $-0.52\pm0.03~{\rm rad~Hz^{-1}}$ and $-0.26\pm0.01~{\rm rad~Hz^{-1}}$ below and above the break, respectively. In this work, we binned the \RXTE measurements in QPO frequency and fitted them with a broken line, obtaining parameters consistent with those reported by \cite{Zhang2020}; the small residual differences most likely stem from not using the same binning scheme. For QPO frequencies above the break, $\alpha$ is negative, meaning the lags are soft, and follows a linear trend with the QPO frequency with a slope of $-0.28\pm0.01~{\rm rad~Hz^{-1}}$. Below the break, $\alpha$ becomes positive, indicating that the lags are hard, and follows a steeper correlation with the QPO frequency with a slope of $-0.53\pm0.03~{\rm rad~Hz^{-1}}$. In contrast, the QPOs detected in the \NICER data show no evidence of a hard-lag regime. In Fig.~\ref{fig:Slopevsnu}, all weighted-average values of $\alpha$ remain negative, indicating soft lags over the full frequency range. Fitting a linear relation between $\alpha$ and the QPO frequency yields a slope of $-0.21 \pm 0.02~{\rm rad~ Hz^{-1}}$, consistent with the \RXTE correlation above the break and significantly different from the steeper relation observed below the break. This suggests that the physical mechanism responsible for the soft lags during the dimming of \grs is the same that was operating in the soft-lag regime of the \RXTE data.

\citet{Zhang2020} find that the QPO fractional rms and the QPO broadband lags are closely related, with the rms amplitude peaking at zero lag and decreasing as the lags become more positive or more negative (their Fig.~8). They further show (their Fig.~6) that this translates into an rms-frequency relation that peaks at $\nu_{\rm QPO}\approx2$~Hz, coincident with the frequency at which the QPO lags change sign. In Fig.~\ref{fig:rmsvsslope}, we present an equivalent figure, plotting the QPO fractional rms as a function of QPO frequency, with the slope of the phase-lag energy spectrum, $\alpha$, encoded in the colour scale. Since negative (positive) values of $\alpha$ correspond to soft (hard) lags, this figure combines the information contained in Figs.~6 and 8 of \citet{Zhang2020}. As shown by \citet{Zhang2020}, the QPO fractional rms increases with QPO frequency while $\alpha$ decreases towards zero, reaches a maximum at $\nu_{\rm QPO}\approx2$~Hz where $\alpha\approx0$, and then decreases again as $\alpha$ becomes increasingly negative at higher frequencies. In our \NICER data, all points show negative-slopes (i.e. soft-lag), with fractional rms values spanning from $\sim7\%$ to $\sim10\%$. These amplitudes are systematically lower than those measured with \RXTE at comparable frequencies, although, as discussed above, this discrepancy may be attributed to the different energy coverage and instrumental responses of the two missions. This offset is substantially reduced after applying the empirical scaling factor derived from the comparison of the RXTE and NICER energy-dependent rms spectra at similar QPO frequencies. While the rms spectra of the two instruments are consistent within the uncertainties at $\sim3$~Hz, they become progressively more discrepant towards lower QPO frequencies, even over the common energy ranges (Fig.~\ref{fig:rmsspectraRXTEvsNICER}). Although this trend is only marginally significant, it may indicate that the QPO fractional rms amplitudes differ intrinsically between the RXTE and NICER epochs at low QPO frequencies. Nevertheless, the \NICER data show no evidence for a turnover such as the one observed in the \RXTE data. This behaviour further supports the interpretation that the radiative properties of the QPO in the \NICER observations are determined by the same mechanism responsible for those properties in the soft-lag regime in the \RXTE observations, and suggests that the mechanism governing the hard-lag regime is not active during the \NICER observations. During the \RXTE observations, the radio emission from the jet was very strong whenever the QPO frequency was below 2~Hz, whereas no radio emission was detected during \NICER data at any frequency. This strongly suggests that the jet is responsible for the hard-lag regime.

 The absence of a hard-lag regime at low QPO frequencies during the dimming of \grs that we find here is consistent with the results of \citet{Koljonen&Hovatta2021} and \citet{Majumder2026}. \citet{Koljonen&Hovatta2021} analysed a \NICER dataset that partially overlaps with ours and find that the broadband QPO lags between the $2-4$~keV and $5-10$~keV energy bands were consistent with zero once the QPO frequency dropped below $\sim2$~Hz, in contrast to the hard lags historically reported with \RXTE.
 \citet{Majumder2026} analysed \textit{AstroSat} observations from March 2016 to March 2019, partially overlapping with our \NICER dataset, and detect type-C QPOs with centroid frequencies between $\sim$ 1.4 and 7.4~Hz. The authors report exclusively soft time lags between the $6-20$~keV and $3-6$~keV energy bands, with no sign reversal near $\sim$2~Hz. Although they only have three observations with QPO frequencies below 2~Hz (see their Table~1), only one of which significantly deviates from the \RXTE trend (see their Fig.~4), their result strengthens the picture emerging from our \NICER analysis, indicating that the absence of hard lags is not instrument-dependent but reflects an intrinsic property of the source during this epoch.

\citet{eijnden2017} investigated the dependence of QPO phase lags upon the QPO frequency in a sample of sources with different inclinations. In their Fig.~3, these authors show that, at high QPO frequencies, the high (low) inclination sources exhibit soft (hard) lags and, as the QPO frequency decreases below $\sim2$~Hz, the phase lags tend to approach values close to zero. A remarkable exception was \grs, which displayed increasingly strong hard lags at low frequencies. The observations of this source included in their analysis were a subsample of the data used by \citet{Zhang2020}. In this context, our \NICER results and those from \citet{Majumder2026} using \textit{AstroSat} data provide an important new perspective on \grs, indicating that, at least during the dimming phase, and when the radio emission is quenched, \grs is no longer an outlier but follows the common low-frequency trend of phase lags seen in the other sources. When probing QPO frequencies below $\sim2$~Hz, we detect only soft lags during the dimming phase, suggesting that the lag behaviour of \grs is not solely determined by the QPO frequency.  This behaviour is consistent with the picture in which \grs, despite its extraordinary phenomenology during outburst, behaves more like a typical BHXB during its transition towards the canonical hard state \citep{Koljonen&Hovatta2021,motta2021}. The hard-lag regime observed in \RXTE data is therefore not an intrinsic property of \grs at low QPO frequencies, but instead depends on other properties of the accreting/outflowing material near the source.

In addition to the timing analysis, we examined the spectral properties of \grs by exploring the evolution of the power-law photon index of the Comptonisation component, $\Gamma$. Since the X-ray spectrum in this state is dominated by the Comptonised component, $\Gamma$ traces the hardness ratio of the source \citep{Garcia2022}. We find a clear correlation between $\Gamma$ and the QPO frequency for the \NICER observations, in which $\Gamma$ increases (the spectrum softens) as the QPO frequency increases \citep[this was previously reported by][]{Zhou2022}. We also compared this relation with \RXTE observations of \grs, including simultaneous radio flux measurements. In the \RXTE data, we distinguish two distinct branches corresponding to low and high radio flux levels. Remarkably, the \NICER measurements align with the \RXTE branch corresponding to weak radio fluxes, consistent with the fact that AMI-LA observations taken almost simultaneously with \NICER show radio fluxes below $\sim5$~mJy. 
This alignment with the low-flux branch reinforces the picture put forward by \citet{Koljonen&Hovatta2021}, who showed that, over the same decay, the mm/X-ray luminosity correlation of \grs shifted from the historically steep, anomalous relation associated with the source's persistently bright outburst to a shallower slope ($L_{\rm radio}\propto L_{\rm X}^{0.6}$) matching the canonical hard-state correlation followed by other BHXBs. These findings provide another indication that, during the dimming phase, \grs was behaving as a regular hard-state BHXB.
%\citet{Majumder2026},  found that the soft lags observed below $\sim2$~Hz are associated with lower Comptonized flux than in the \RXTE hard-lag regime, and suggested that this may be linked to a higher coronal optical depth. 

\cite{Karpouzas2021} and \cite{Garcia2022} fitted the energy-dependent phase-lag and fractional rms amplitude spectra of the type-C QPO data of \grs using a variable-Comptonisation model \citep[][see \citealt{Garcia2021} and \citealt{bellavita2022} for updated versions of the model]{Karpouzas2020} that includes feedback, whereby a fraction of the up-scattered photons impinge back onto the disc. They find that when the QPO frequency is above $\sim2$~Hz, the feedback parameter is high, indicating that a large fraction of the Comptonised emission is reprocessed by the disc, which naturally explains the soft lags. Instead, for QPO frequencies below $\sim 2$~Hz, they find low feedback values, which suggests either a corona that does not cover the inner parts of the disc, one that is less vertically extended and therefore illuminates a smaller fraction of the disc surface, or one in which part of the coronal plasma is being redirected into an outflow or jet, reducing the fraction of photons that return to the disc. The fact that the \NICER data show exclusively soft QPO lags, together with $\Gamma$ values consistent with the low-radio-flux branch in the plot of $\Gamma$ vs. QPO frequency, suggests that in \grs the configuration of the corona during the dimming phase of the outburst is analogous to that of the soft-lag solutions identified above $\sim2$~Hz in the \RXTE data. In this picture, the corona is compact enough and sufficiently coupled to the disc, to sustain strong feedback. 

Although $kT_{\rm e}$ cannot be well constrained in our \NICER fits, joint \NICER--HXMT analyses during the decay phase found coronal temperatures reaching up to $\sim150$ keV \citep{Zhou2025}. By fitting the variable-Comptonisation model, \citet{Karpouzas2021} find that the coronal temperature in the \RXTE data is systematically higher when the QPO frequency is above $\sim2$~Hz than when it is below $\sim2$~Hz. The high temperatures during the \NICER era are consistent with those associated with the soft-lag regime in the \RXTE data, further supporting the idea that both correspond to similar physical coronal configurations.

\cite{Mendez2022} show that the change of sign of the QPO lags at $\sim2$~Hz in the \RXTE data marks a sharp transition in coronal geometry coincident with the onset of strong radio activity. \cite{Garcia2022} find that below $\sim2$~Hz the corona size shrinks as the QPO frequency increases, while above $\sim2$~Hz, the corona expands again with frequency. \cite{Mendez2022} and \cite{Garcia2022} propose that when the QPO frequency drops below $\sim2$~Hz the corona expands vertically, the feedback fraction decreases, and part of the coronal plasma is redirected into the jet, giving rise to hard lags and radio flares. In this picture, the hard lags arise naturally from Comptonisation in a corona with low feedback, and the jet is a concurrent manifestation of the same transition in the inflow--outflow configuration. Instead, above $\sim2$~Hz, the corona retains a disc-coupled geometry and the jet is weak or quenched, in agreement with the low radio fluxes observed simultaneously. The absence of hard lags and of a turnover in the rms–lag relation in the \NICER data therefore implies that such a transition in the coronal geometry is not reached during the dimming phase.

Our results indicate that the sign and magnitude of the QPO phase lags in \grs are not uniquely determined by the QPO frequency, as previously assumed, but hinge also on the accretion state and the presence of a jet. This highlights the importance of considering the coupling between inflow and outflow when interpreting QPO timing properties. The dimming phase observed with \NICER therefore provides a unique window to isolate the soft-lag regime and study the corona in the absence of strong jet activity. Our results reinforce the idea that the QPO phase lags provide a sensitive tracer of the coronal geometry and its coupling to the jet in \grs.

\begin{acknowledgements}
We thank the referee for the insightful suggestions that helped us improve this paper. CB is a CONICET fellow. CB acknowledges the financial support provided by the Leids Kerkhoven Bosscha Fonds (LKBF) and the Kapteyn Astronomical Institute to attend conferences. FG is a CONICET researcher. PY acknowledges support from the China Scholarship Council (CSC), No. 202304910059. This research was supported by the International Space Science Institute (ISSI) in Bern, through ISSI International Team project
$\#486$. CB, FG, MM and PY thank the Lorentz Center workshop: Compact Objects in 3D -- Steps towards X-ray Polarimetric-Spectral-Timing for the insightful discussions that contributed to the development of this project. This research has made use of NASA's Astrophysics Data System Bibliographic Services.
\end{acknowledgements}

\bibliographystyle{bibtex/aa}
\bibliography{aanda} 

@ARTICLE{Zhang2020,
       author = {{Zhang}, Liang and {M{\'e}ndez}, Mariano and {Altamirano}, Diego and {Qu}, Jinlu and {Chen}, Li and {Karpouzas}, Konstantinos and {Belloni}, Tomaso M. and {Bu}, Qingcui and {Huang}, Yue and {Ma}, Xiang and {Tao}, Lian and {Wang}, Yanan},
        title = "{A systematic analysis of the phase lags associated with the type-C quasi-periodic oscillation in GRS 1915+105}",
      journal = {\mnras},
         year = 2020,
        month = may,
       volume = {494},
       number = {1},
        pages = {1375-1386},
          doi = {10.1093/mnras/staa797},
archivePrefix = {arXiv},
       eprint = {2003.08928},
 primaryClass = {astro-ph.HE},
       adsurl = {https://ui.adsabs.harvard.edu/abs/2020MNRAS.494.1375Z}
}

@ARTICLE{Koljonen&Hovatta2021,
       author = {{Koljonen}, K.~I.~I. and {Hovatta}, T.},
        title = "{ALMA/NICER observations of GRS 1915+105 indicate a return to a hard state}",
      journal = {\aap},
         year = 2021,
        month = mar,
       volume = {647},
          eid = {A173},
        pages = {A173},
          doi = {10.1051/0004-6361/202039581},
archivePrefix = {arXiv},
       eprint = {2102.00693},
 primaryClass = {astro-ph.HE},
       adsurl = {https://ui.adsabs.harvard.edu/abs/2021A&A...647A.173K}
}

@ARTICLE{reig2003,
       author = {{Reig}, P. and {Kylafis}, N.~D. and {Giannios}, D.},
        title = "{Energy and time-lag spectra of galactic black-hole X-ray sources in the low/hard state}",
      journal = {\aap},
         year = 2003,
        month = may,
       volume = {403},
        pages = {L15-L18},
          doi = {10.1051/0004-6361:20030449},
archivePrefix = {arXiv},
       eprint = {astro-ph/0303585},
 primaryClass = {astro-ph},
       adsurl = {https://ui.adsabs.harvard.edu/abs/2003A&A...403L..15R}
}

@ARTICLE{Giannios2004,
       author = {{Giannios}, D. and {Kylafis}, N.~D. and {Psaltis}, D.},
        title = "{Spectra and time variability of Galactic black-hole X-ray sources in the low/hard state}",
      journal = {\aap},
         year = 2004,
        month = oct,
       volume = {425},
        pages = {163-169},
          doi = {10.1051/0004-6361:20041002},
archivePrefix = {arXiv},
       eprint = {astro-ph/0405569},
 primaryClass = {astro-ph},
       adsurl = {https://ui.adsabs.harvard.edu/abs/2004A&A...425..163G}
}

@ARTICLE{Jin2026,
       author = {{Jin}, Pei and {M{\'e}ndez}, Mariano and {Garc{\'\i}a}, Federico and {Altamirano}, Diego and {Vincentelli}, Federico M.},
        title = "{Black-hole X-ray binary Swift J1727.8─1613 shows simultaneous Type-B and Type-C quasiperiodic oscillations across the hard-intermediate and soft-intermediate states}",
      journal = {\aap},
         year = 2026,
        month = feb,
       volume = {706},
          eid = {A208},
        pages = {A208},
          doi = {10.1051/0004-6361/202555907},
archivePrefix = {arXiv},
       eprint = {2510.10353},
 primaryClass = {astro-ph.HE},
       adsurl = {https://ui.adsabs.harvard.edu/abs/2026A&A...706A.208J}
}

@ARTICLE{castrotirado1992,
       author = {{Castro-Tirado}, A.~J. and {Brandt}, S. and {Lund}, N.},
        title = "{GRS 1915+105}",
      journal = {\iaucirc},
         year = 1992,
        month = aug,
       volume = {5590},
        pages = {2},
       adsurl = {https://ui.adsabs.harvard.edu/abs/1992IAUC.5590....2C}
}

@ARTICLE{fender2025,
       author = {{Fender}, R.~P. and {Motta}, S.~E.},
        title = "{The connection between the fastest astrophysical jets and the spin axis of their black hole}",
      journal = {Nature Astronomy},
         year = 2025,
        month = dec,
       volume = {9},
        pages = {1854-1859},
          doi = {10.1038/s41550-025-02665-w},
       adsurl = {https://ui.adsabs.harvard.edu/abs/2025NatAs...9.1854F}
}

@ARTICLE{Pooley1997,
       author = {{Pooley}, G.~G. and {Fender}, R.~P.},
        title = "{The variable radio emission from GRS 1915+=105}",
      journal = {\mnras},
         year = 1997,
        month = dec,
       volume = {292},
       number = {4},
        pages = {925-933},
          doi = {10.1093/mnras/292.4.925},
archivePrefix = {arXiv},
       eprint = {astro-ph/9708171},
 primaryClass = {astro-ph},
       adsurl = {https://ui.adsabs.harvard.edu/abs/1997MNRAS.292..925P}
}

@ARTICLE{Negoro2018,
       author = {{Negoro}, H. and {Tachibana}, Y. and {Kawai}, N. and {Yamaoka}, K. and {Ueda}, Y. and {Nakajima}, M. and {Sakamaki}, A. and {Maruyama}, W. and {Mihara}, T. and {Nakahira}, S. and {Yatabe}, F. and {Takao}, Y. and {Matsuoka}, M. and {Sakamoto}, T. and {Serino}, M. and {Sugita}, S. and {Kawakubo}, Y. and {Hashimoto}, T. and {Yoshida}, A. and {Sugizaki}, M. and {Morita}, K. and {Ueno}, S. and {Tomida}, H. and {Ishikawa}, M. and {Sugawara}, Y. and {Isobe}, N. and {Shimomukai}, R. and {Tanimoto}, A. and {Morita}, T. and {Yamada}, S. and {Tsuboi}, Y. and {Iwakiri}, W. and {Sasaki}, R. and {Kawai}, H. and {Sato}, T. and {Tsunemi}, H. and {Yoneyama}, T. and {Yamauchi}, M. and {Hidaka}, K. and {Iwahori}, S. and {Kawamuro}, T. and {Shidatsu}, M.},
        title = "{MAXI/GSC observes GRS 1915+105 in the X-ray faintest state in the last 22 years}",
      journal = {The Astronomer's Telegram},
         year = 2018,
        month = jul,
       volume = {11828},
        pages = {1},
       adsurl = {https://ui.adsabs.harvard.edu/abs/2018ATel11828....1N}
}

@ARTICLE{belloni2000,
       author = {{Belloni}, T. and {Klein-Wolt}, M. and {M{\'e}ndez}, M. and {van der Klis}, M. and {van Paradijs}, J.},
        title = "{A model-independent analysis of the variability of GRS 1915+105}",
      journal = {\aap},
         year = 2000,
        month = mar,
       volume = {355},
        pages = {271-290},
          doi = {10.48550/arXiv.astro-ph/0001103},
archivePrefix = {arXiv},
       eprint = {astro-ph/0001103},
 primaryClass = {astro-ph},
       adsurl = {https://ui.adsabs.harvard.edu/abs/2000A&A...355..271B}
}

@ARTICLE{Garcia2021,
       author = {{Garc{\'\i}a}, Federico and {M{\'e}ndez}, Mariano and {Karpouzas}, Konstantinos and {Belloni}, Tomaso and {Zhang}, Liang and {Altamirano}, Diego},
        title = "{A two-component Comptonization model for the type-B QPO in MAXI J1348-630}",
      journal = {\mnras},
         year = 2021,
        month = mar,
       volume = {501},
       number = {3},
        pages = {3173-3182},
          doi = {10.1093/mnras/staa3944},
archivePrefix = {arXiv},
       eprint = {2012.10354},
 primaryClass = {astro-ph.HE},
       adsurl = {https://ui.adsabs.harvard.edu/abs/2021MNRAS.501.3173G}
}

@ARTICLE{Tetarenko2016,
       author = {{Tetarenko}, B.~E. and {Sivakoff}, G.~R. and {Heinke}, C.~O. and {Gladstone}, J.~C.},
        title = "{WATCHDOG: A Comprehensive All-sky Database of Galactic Black Hole X-ray Binaries}",
      journal = {\apjs},
         year = 2016,
        month = feb,
       volume = {222},
       number = {2},
          eid = {15},
        pages = {15},
          doi = {10.3847/0067-0049/222/2/15},
archivePrefix = {arXiv},
       eprint = {1512.00778},
 primaryClass = {astro-ph.HE},
       adsurl = {https://ui.adsabs.harvard.edu/abs/2016ApJS..222...15T}
}

@ARTICLE{fender2009,
       author = {{Fender}, R.~P. and {Homan}, J. and {Belloni}, T.~M.},
        title = "{Jets from black hole X-ray binaries: testing, refining and extending empirical models for the coupling to X-rays}",
      journal = {\mnras},
         year = 2009,
        month = jul,
       volume = {396},
       number = {3},
        pages = {1370-1382},
          doi = {10.1111/j.1365-2966.2009.14841.x},
archivePrefix = {arXiv},
       eprint = {0903.5166},
 primaryClass = {astro-ph.HE},
       adsurl = {https://ui.adsabs.harvard.edu/abs/2009MNRAS.396.1370F}
}

@ARTICLE{Motta2016,
       author = {{Motta}, S.~E.},
        title = "{Quasi periodic oscillations in black hole binaries}",
      journal = {Astronomische Nachrichten},
         year = 2016,
        month = may,
       volume = {337},
       number = {4-5},
        pages = {398},
          doi = {10.1002/asna.201612320},
archivePrefix = {arXiv},
       eprint = {1603.07885},
 primaryClass = {astro-ph.HE},
       adsurl = {https://ui.adsabs.harvard.edu/abs/2016AN....337..398M}
}

@ARTICLE{uttley2015,
       author = {{Uttley}, Philip and {Klein-Wolt}, Marc},
        title = "{The remarkable timing properties of a `hypersoft' state in GRO J1655-40}",
      journal = {\mnras},
         year = 2015,
        month = jul,
       volume = {451},
       number = {1},
        pages = {475-485},
          doi = {10.1093/mnras/stv978},
archivePrefix = {arXiv},
       eprint = {1504.08313},
 primaryClass = {astro-ph.HE},
       adsurl = {https://ui.adsabs.harvard.edu/abs/2015MNRAS.451..475U}
}

@ARTICLE{dunn2010,
       author = {{Dunn}, R.~J.~H. and {Fender}, R.~P. and {K{\"o}rding}, E.~G. and {Belloni}, T. and {Cabanac}, C.},
        title = "{A global spectral study of black hole X-ray binaries}",
      journal = {\mnras},
         year = 2010,
        month = mar,
       volume = {403},
       number = {1},
        pages = {61-82},
          doi = {10.1111/j.1365-2966.2010.16114.x},
archivePrefix = {arXiv},
       eprint = {0912.0142},
 primaryClass = {astro-ph.HE},
       adsurl = {https://ui.adsabs.harvard.edu/abs/2010MNRAS.403...61D}
}

@ARTICLE{munozdarias2011,
       author = {{Mu{\~n}oz-Darias}, T. and {Motta}, S. and {Belloni}, T.~M.},
        title = "{Fast variability as a tracer of accretion regimes in black hole transients}",
      journal = {\mnras},
         year = 2011,
        month = jan,
       volume = {410},
       number = {1},
        pages = {679-684},
          doi = {10.1111/j.1365-2966.2010.17476.x},
archivePrefix = {arXiv},
       eprint = {1008.0558},
 primaryClass = {astro-ph.HE},
       adsurl = {https://ui.adsabs.harvard.edu/abs/2011MNRAS.410..679M}
}

@ARTICLE{rushton2010,
       author = {{Rushton}, A. and {Spencer}, R. and {Fender}, R. and {Pooley}, G.},
        title = "{Steady jets from radiatively efficient hard states in GRS 1915+105}",
      journal = {\aap},
         year = 2010,
        month = dec,
       volume = {524},
          eid = {A29},
        pages = {A29},
          doi = {10.1051/0004-6361/201014929},
archivePrefix = {arXiv},
       eprint = {1101.4945},
 primaryClass = {astro-ph.HE},
       adsurl = {https://ui.adsabs.harvard.edu/abs/2010A&A...524A..29R}
}

@ARTICLE{Mirabel1994,
       author = {{Mirabel}, I.~F. and {Rodr{\'\i}guez}, L.~F.},
        title = "{A superluminal source in the Galaxy}",
      journal = {\nat},
         year = 1994,
        month = sep,
       volume = {371},
       number = {6492},
        pages = {46-48},
          doi = {10.1038/371046a0},
       adsurl = {https://ui.adsabs.harvard.edu/abs/1994Natur.371...46M}
}

@ARTICLE{castrotirado1994,
       author = {{Castro-Tirado}, Alberto J. and {Brandt}, Soren and {Lund}, Niels and {Lapshov}, Igor and {Sunyaev}, Rashid A. and {Shlyapnikov}, Aleksei A. and {Guziy}, Sergei and {Pavlenko}, Elena P.},
        title = "{Discovery and Observations by WATCH of the X-Ray Transient GRS 1915+105}",
      journal = {\apjs},
         year = 1994,
        month = jun,
       volume = {92},
        pages = {469},
          doi = {10.1086/191998},
       adsurl = {https://ui.adsabs.harvard.edu/abs/1994ApJS...92..469C}
}

@ARTICLE{vanderklis1994,
       author = {{van der Klis}, M.},
        title = "{Similarities in Neutron Star and Black Hole Accretion}",
      journal = {\apjs},
         year = 1994,
        month = jun,
       volume = {92},
        pages = {511},
          doi = {10.1086/192006},
       adsurl = {https://ui.adsabs.harvard.edu/abs/1994ApJS...92..511V}
}

@INPROCEEDINGS{Belloni2016,
       author = {{Belloni}, Tomaso M. and {Motta}, Sara E.},
        title = "{Transient Black Hole Binaries}",
    booktitle = {Astrophysics of Black Holes: From Fundamental Aspects to Latest Developments},
         year = 2016,
       editor = {{Bambi}, Cosimo},
       series = {Astrophysics and Space Science Library},
       volume = {440},
        month = jan,
        pages = {61},
          doi = {10.1007/978-3-662-52859-4_2},
archivePrefix = {arXiv},
       eprint = {1603.07872},
 primaryClass = {astro-ph.HE},
       adsurl = {https://ui.adsabs.harvard.edu/abs/2016ASSL..440...61B}
}

@ARTICLE{Zhou2025,
       author = {{Zhou}, M. and {Grinberg}, V. and {Santangelo}, A. and {Bambi}, C. and {Bu}, Q. and {Diez}, C.~M. and {Kong}, L. and {Steiner}, J.~F. and {Tuo}, Y.},
        title = "{Dimming GRS 1915+105 observed with NICER and Insight{\textendash}HXMT}",
      journal = {\aap},
         year = 2025,
        month = feb,
       volume = {694},
          eid = {A104},
        pages = {A104},
          doi = {10.1051/0004-6361/202451558},
archivePrefix = {arXiv},
       eprint = {2501.03804},
 primaryClass = {astro-ph.HE},
       adsurl = {https://ui.adsabs.harvard.edu/abs/2025A&A...694A.104Z}
}

@ARTICLE{Belloni2024,
       author = {{Belloni}, Tomaso M. and {M{\'e}ndez}, Mariano and {Garc{\'\i}a}, Federico and {Bhattacharya}, Dipankar},
        title = "{Fast-varying time lags in the quasi-periodic oscillation in GRS 1915 + 105}",
      journal = {\mnras},
         year = 2024,
        month = jan,
       volume = {527},
       number = {3},
        pages = {7136-7143},
          doi = {10.1093/mnras/stad3639},
archivePrefix = {arXiv},
       eprint = {2311.13467},
 primaryClass = {astro-ph.HE},
       adsurl = {https://ui.adsabs.harvard.edu/abs/2024MNRAS.527.7136B}
}

@ARTICLE{Mendez2024,
       author = {{M{\'e}ndez}, Mariano and {Peirano}, Valentina and {Garc{\'\i}a}, Federico and {Belloni}, Tomaso and {Altamirano}, Diego and {Alabarta}, Kevin},
        title = "{Unveiling hidden variability components in accreting X-ray binaries using both the Fourier power and cross-spectra}",
      journal = {\mnras},
         year = 2024,
        month = jan,
       volume = {527},
       number = {3},
        pages = {9405-9430},
          doi = {10.1093/mnras/stad3786},
archivePrefix = {arXiv},
       eprint = {2312.03476},
 primaryClass = {astro-ph.HE},
       adsurl = {https://ui.adsabs.harvard.edu/abs/2024MNRAS.527.9405M}
}

@INPROCEEDINGS{Gendreau2016,
       author = {{Gendreau}, Keith C. and {Arzoumanian}, Zaven and {Adkins}, Phillip W. and {Albert}, Cheryl L. and {Anders}, John F. and {Aylward}, Andrew T. and {Baker}, Charles L. and {Balsamo}, Erin R. and {Bamford}, William A. and {Benegalrao}, Suyog S. and {Berry}, Daniel L. and {Bhalwani}, Shiraz and {Black}, J. Kevin and {Blaurock}, Carl and {Bronke}, Ginger M. and {Brown}, Gary L. and {Budinoff}, Jason G. and {Cantwell}, Jeffrey D. and {Cazeau}, Thoniel and {Chen}, Philip T. and {Clement}, Thomas G. and {Colangelo}, Andrew T. and {Coleman}, Jerry S. and {Coopersmith}, Jonathan D. and {Dehaven}, William E. and {Doty}, John P. and {Egan}, Mark D. and {Enoto}, Teruaki and {Fan}, Terry W. and {Ferro}, Deneen M. and {Foster}, Richard and {Galassi}, Nicholas M. and {Gallo}, Luis D. and {Green}, Chris M. and {Grosh}, Dave and {Ha}, Kong Q. and {Hasouneh}, Monther A. and {Heefner}, Kristofer B. and {Hestnes}, Phyllis and {Hoge}, Lisa J. and {Jacobs}, Tawanda M. and {J{\o}rgensen}, John L. and {Kaiser}, Michael A. and {Kellogg}, James W. and {Kenyon}, Steven J. and {Koenecke}, Richard G. and {Kozon}, Robert P. and {LaMarr}, Beverly and {Lambertson}, Mike D. and {Larson}, Anne M. and {Lentine}, Steven and {Lewis}, Jesse H. and {Lilly}, Michael G. and {Liu}, Kuochia Alice and {Malonis}, Andrew and {Manthripragada}, Sridhar S. and {Markwardt}, Craig B. and {Matonak}, Bryan D. and {Mcginnis}, Isaac E. and {Miller}, Roger L. and {Mitchell}, Alissa L. and {Mitchell}, Jason W. and {Mohammed}, Jelila S. and {Monroe}, Charles A. and {Montt de Garcia}, Kristina M. and {Mul{\'e}}, Peter D. and {Nagao}, Louis T. and {Ngo}, Son N. and {Norris}, Eric D. and {Norwood}, Dwight A. and {Novotka}, Joseph and {Okajima}, Takashi and {Olsen}, Lawrence G. and {Onyeachu}, Chimaobi O. and {Orosco}, Henry Y. and {Peterson}, Jacqualine R. and {Pevear}, Kristina N. and {Pham}, Karen K. and {Pollard}, Sue E. and {Pope}, John S. and {Powers}, Daniel F. and {Powers}, Charles E. and {Price}, Samuel R. and {Prigozhin}, Gregory Y. and {Ramirez}, Julian B. and {Reid}, Winston J. and {Remillard}, Ronald A. and {Rogstad}, Eric M. and {Rosecrans}, Glenn P. and {Rowe}, John N. and {Sager}, Jennifer A. and {Sanders}, Claude A. and {Savadkin}, Bruce and {Saylor}, Maxine R. and {Schaeffer}, Alexander F. and {Schweiss}, Nancy S. and {Semper}, Sean R. and {Serlemitsos}, Peter J. and {Shackelford}, Larry V. and {Soong}, Yang and {Struebel}, Jonathan and {Vezie}, Michael L. and {Villasenor}, Joel S. and {Winternitz}, Luke B. and {Wofford}, George I. and {Wright}, Michael R. and {Yang}, Mike Y. and {Yu}, Wayne H.},
        title = "{The Neutron star Interior Composition Explorer (NICER): design and development}",
    booktitle = {Space Telescopes and Instrumentation 2016: Ultraviolet to Gamma Ray},
         year = 2016,
       editor = {{den Herder}, Jan-Willem A. and {Takahashi}, Tadayuki and {Bautz}, Marshall},
       series = {Society of Photo-Optical Instrumentation Engineers (SPIE) Conference Series},
       volume = {9905},
        month = jul,
          eid = {99051H},
        pages = {99051H},
          doi = {10.1117/12.2231304},
       adsurl = {https://ui.adsabs.harvard.edu/abs/2016SPIE.9905E..1HG}
}

@ARTICLE{Motta2011,
       author = {{Motta}, S. and {Mu{\~n}oz-Darias}, T. and {Casella}, P. and {Belloni}, T. and {Homan}, J.},
        title = "{Low-frequency oscillations in black holes: a spectral-timing approach to the case of GX 339-4}",
      journal = {\mnras},
         year = 2011,
        month = dec,
       volume = {418},
       number = {4},
        pages = {2292-2307},
          doi = {10.1111/j.1365-2966.2011.19566.x},
archivePrefix = {arXiv},
       eprint = {1108.0540},
 primaryClass = {astro-ph.HE},
       adsurl = {https://ui.adsabs.harvard.edu/abs/2011MNRAS.418.2292M}
}

@ARTICLE{Casella2005,
       author = {{Casella}, P. and {Belloni}, T. and {Stella}, L.},
        title = "{The ABC of Low-Frequency Quasi-periodic Oscillations in Black Hole Candidates: Analogies with Z Sources}",
      journal = {\apj},
         year = 2005,
        month = aug,
       volume = {629},
       number = {1},
        pages = {403-407},
          doi = {10.1086/431174},
archivePrefix = {arXiv},
       eprint = {astro-ph/0504318},
 primaryClass = {astro-ph},
       adsurl = {https://ui.adsabs.harvard.edu/abs/2005ApJ...629..403C}
}

@ARTICLE{Casella2004,
       author = {{Casella}, P. and {Belloni}, T. and {Homan}, J. and {Stella}, L.},
        title = "{A study of the low-frequency quasi-periodic oscillations in the X-ray light curves of the black hole candidate <ASTROBJ>XTE J1859+226</ASTROBJ>}",
      journal = {\aap},
         year = 2004,
        month = nov,
       volume = {426},
        pages = {587-600},
          doi = {10.1051/0004-6361:20041231},
archivePrefix = {arXiv},
       eprint = {astro-ph/0407262},
 primaryClass = {astro-ph},
       adsurl = {https://ui.adsabs.harvard.edu/abs/2004A&A...426..587C}
}

@ARTICLE{Mendez2022,
       author = {{M{\'e}ndez}, Mariano and {Karpouzas}, Konstantinos and {Garc{\'\i}a}, Federico and {Zhang}, Liang and {Zhang}, Yuexin and {Belloni}, Tomaso M. and {Altamirano}, Diego},
        title = "{Coupling between the accreting corona and the relativistic jet in the microquasar GRS 1915+105}",
      journal = {Nature Astronomy},
         year = 2022,
        month = mar,
       volume = {6},
        pages = {577-583},
          doi = {10.1038/s41550-022-01617-y},
archivePrefix = {arXiv},
       eprint = {2203.02963},
 primaryClass = {astro-ph.HE},
       adsurl = {https://ui.adsabs.harvard.edu/abs/2022NatAs...6..577M}
}

@ARTICLE{Garcia2022,
       author = {{Garc{\'\i}a}, Federico and {Karpouzas}, Konstantinos and {M{\'e}ndez}, Mariano and {Zhang}, Liang and {Zhang}, Yuexin and {Belloni}, Tomaso and {Altamirano}, Diego},
        title = "{The evolving properties of the corona of GRS 1915+105: a spectral-timing perspective through variable-Comptonization modelling}",
      journal = {\mnras},
         year = 2022,
        month = jul,
       volume = {513},
       number = {3},
        pages = {4196-4207},
          doi = {10.1093/mnras/stac1202},
archivePrefix = {arXiv},
       eprint = {2204.13279},
 primaryClass = {astro-ph.HE},
       adsurl = {https://ui.adsabs.harvard.edu/abs/2022MNRAS.513.4196G}
}

@ARTICLE{Mendez1997,
       author = {{M{\'e}ndez}, Mariano and {van der Klis}, Michiel},
        title = "{The EXOSAT Data on GX 339-4: Further Evidence for an ``Intermediate'' State}",
      journal = {\apj},
         year = 1997,
        month = apr,
       volume = {479},
       number = {2},
        pages = {926-932},
          doi = {10.1086/303914},
archivePrefix = {arXiv},
       eprint = {astro-ph/9611015},
 primaryClass = {astro-ph},
       adsurl = {https://ui.adsabs.harvard.edu/abs/1997ApJ...479..926M}
}

@ARTICLE{Homan2001,
       author = {{Homan}, Jeroen and {Wijnands}, Rudy and {van der Klis}, Michiel and {Belloni}, Tomaso and {van Paradijs}, Jan and {Klein-Wolt}, Marc and {Fender}, Rob and {M{\'e}ndez}, Mariano},
        title = "{Correlated X-Ray Spectral and Timing Behavior of the Black Hole Candidate XTE J1550-564: A New Interpretation of Black Hole States}",
      journal = {\apjs},
         year = 2001,
        month = feb,
       volume = {132},
       number = {2},
        pages = {377-402},
          doi = {10.1086/318954},
archivePrefix = {arXiv},
       eprint = {astro-ph/0001163},
 primaryClass = {astro-ph},
       adsurl = {https://ui.adsabs.harvard.edu/abs/2001ApJS..132..377H}
}

@ARTICLE{Belloni2005,
       author = {{Belloni}, T. and {Homan}, J. and {Casella}, P. and {van der Klis}, M. and {Nespoli}, E. and {Lewin}, W.~H.~G. and {Miller}, J.~M. and {M{\'e}ndez}, M.},
        title = "{The evolution of the timing properties of the black-hole transient GX 339-4 during its 2002/2003 outburst}",
      journal = {\aap},
         year = 2005,
        month = sep,
       volume = {440},
       number = {1},
        pages = {207-222},
          doi = {10.1051/0004-6361:20042457},
archivePrefix = {arXiv},
       eprint = {astro-ph/0504577},
 primaryClass = {astro-ph},
       adsurl = {https://ui.adsabs.harvard.edu/abs/2005A&A...440..207B}
}

@ARTICLE{Mitsuda1984,
       author = {{Mitsuda}, K. and {Inoue}, H. and {Koyama}, K. and {Makishima}, K. and {Matsuoka}, M. and {Ogawara}, Y. and {Shibazaki}, N. and {Suzuki}, K. and {Tanaka}, Y. and {Hirano}, T.},
        title = "{Energy Spectra of Low-Mass Binary X-Ray Sources Observed from Tenma}",
      journal = {\pasj},
         year = 1984,
        month = dec,
       volume = {36},
       number = {4},
        pages = {741-759},
          doi = {10.1093/pasj/36.4.741},
       adsurl = {https://ui.adsabs.harvard.edu/abs/1984PASJ...36..741M}
}

@ARTICLE{Makishima1986,
       author = {{Makishima}, K. and {Maejima}, Y. and {Mitsuda}, K. and {Bradt}, H.~V. and {Remillard}, R.~A. and {Tuohy}, I.~R. and {Hoshi}, R. and {Nakagawa}, M.},
        title = "{Simultaneous X-Ray and Optical Observations of GX 339-4 in an X-Ray High State}",
      journal = {\apj},
         year = 1986,
        month = sep,
       volume = {308},
        pages = {635},
          doi = {10.1086/164534},
       adsurl = {https://ui.adsabs.harvard.edu/abs/1986ApJ...308..635M}
}

@ARTICLE{Zycki1999,
       author = {{{\.Z}ycki}, Piotr T. and {Done}, Chris and {Smith}, David A.},
        title = "{The 1989 May outburst of the soft X-ray transient GS 2023+338 (V404 Cyg)}",
      journal = {\mnras},
         year = 1999,
        month = nov,
       volume = {309},
       number = {3},
        pages = {561-575},
          doi = {10.1046/j.1365-8711.1999.02885.x},
archivePrefix = {arXiv},
       eprint = {astro-ph/9904304},
 primaryClass = {astro-ph},
       adsurl = {https://ui.adsabs.harvard.edu/abs/1999MNRAS.309..561Z}
}

@ARTICLE{Zdziarski1996,
       author = {{Zdziarski}, A.~A. and {Johnson}, W.~N. and {Magdziarz}, P.},
        title = "{Broad-band {\ensuremath{\gamma}}-ray and X-ray spectra of NGC 4151 and their implications for physical processes and geometry.}",
      journal = {\mnras},
         year = 1996,
        month = nov,
       volume = {283},
       number = {1},
        pages = {193-206},
          doi = {10.1093/mnras/283.1.193},
archivePrefix = {arXiv},
       eprint = {astro-ph/9607015},
 primaryClass = {astro-ph},
       adsurl = {https://ui.adsabs.harvard.edu/abs/1996MNRAS.283..193Z}
}

@ARTICLE{Wilms2000,
       author = {{Wilms}, J. and {Allen}, A. and {McCray}, R.},
        title = "{On the Absorption of X-Rays in the Interstellar Medium}",
      journal = {\apj},
         year = 2000,
        month = oct,
       volume = {542},
       number = {2},
        pages = {914-924},
          doi = {10.1086/317016},
archivePrefix = {arXiv},
       eprint = {astro-ph/0008425},
 primaryClass = {astro-ph},
       adsurl = {https://ui.adsabs.harvard.edu/abs/2000ApJ...542..914W}
}

@ARTICLE{Majumder2026,
       author = {{Majumder}, Prajjwal and {Dutta}, Broja G. and {Nandi}, Anuj},
        title = "{Time-Lag properties associated with LFQPO in X-ray variability classes of GRS 1915+105: Findings from AstroSat}",
      journal = {\mnras},
         year = 2026,
        month = mar,
          doi = {10.1093/mnras/stag556},
archivePrefix = {arXiv},
       eprint = {2603.16305},
 primaryClass = {astro-ph.HE},
       adsurl = {https://ui.adsabs.harvard.edu/abs/2026MNRAS.tmp..525M}
}

@ARTICLE{eijnden2016,
       author = {{van den Eijnden}, Jakob and {Ingram}, Adam and {Uttley}, Phil},
        title = "{Probing the origin of quasi-periodic oscillations: the short-time-scale evolution of phase lags in GRS 1915+105}",
      journal = {\mnras},
         year = 2016,
        month = jun,
       volume = {458},
       number = {4},
        pages = {3655-3666},
          doi = {10.1093/mnras/stw610},
archivePrefix = {arXiv},
       eprint = {1603.03392},
 primaryClass = {astro-ph.HE},
       adsurl = {https://ui.adsabs.harvard.edu/abs/2016MNRAS.458.3655V}
}

@ARTICLE{demarco2015,
       author = {{De Marco}, B. and {Ponti}, G. and {Mu{\~n}oz-Darias}, T. and {Nandra}, K.},
        title = "{Tracing the Reverberation Lag in the Hard State of Black Hole X-Ray Binaries}",
      journal = {\apj},
         year = 2015,
        month = nov,
       volume = {814},
       number = {1},
          eid = {50},
        pages = {50},
          doi = {10.1088/0004-637X/814/1/50},
archivePrefix = {arXiv},
       eprint = {1510.02798},
 primaryClass = {astro-ph.HE},
       adsurl = {https://ui.adsabs.harvard.edu/abs/2015ApJ...814...50D}
}

@ARTICLE{Lee2001,
       author = {{Lee}, Hyong C. and {Misra}, R. and {Taam}, Ronald E.},
        title = "{A Compton Upscattering Model for Soft Lags in the Lower Kilohertz Quasi-periodic Oscillation in 4U 1608-52}",
      journal = {\apjl},
         year = 2001,
        month = mar,
       volume = {549},
       number = {2},
        pages = {L229-L232},
          doi = {10.1086/319171},
archivePrefix = {arXiv},
       eprint = {astro-ph/0102209},
 primaryClass = {astro-ph},
       adsurl = {https://ui.adsabs.harvard.edu/abs/2001ApJ...549L.229L}
}

@ARTICLE{Verner1996,
       author = {{Verner}, D.~A. and {Ferland}, G.~J. and {Korista}, K.~T. and {Yakovlev}, D.~G.},
        title = "{Atomic Data for Astrophysics. II. New Analytic Fits for Photoionization Cross Sections of Atoms and Ions}",
      journal = {\apj},
         year = 1996,
        month = jul,
       volume = {465},
        pages = {487},
          doi = {10.1086/177435},
archivePrefix = {arXiv},
       eprint = {astro-ph/9601009},
 primaryClass = {astro-ph},
       adsurl = {https://ui.adsabs.harvard.edu/abs/1996ApJ...465..487V}
}

@ARTICLE{neilsen2020,
       author = {{Neilsen}, J. and {Homan}, J. and {Steiner}, J.~F. and {Marcel}, G. and {Cackett}, E. and {Remillard}, R.~A. and {Gendreau}, K.},
        title = "{A NICER View of a Highly Absorbed Flare in GRS 1915+105}",
      journal = {\apj},
         year = 2020,
        month = oct,
       volume = {902},
       number = {2},
          eid = {152},
        pages = {152},
          doi = {10.3847/1538-4357/abb598},
archivePrefix = {arXiv},
       eprint = {2010.14512},
 primaryClass = {astro-ph.HE},
       adsurl = {https://ui.adsabs.harvard.edu/abs/2020ApJ...902..152N}
}

@ARTICLE{Motta2019,
       author = {{Motta}, Sara and {Williams}, David and {Fender}, Rob and {Titterington}, David and {Green}, Dave and {Perrott}, Yvette},
        title = "{AMI-LA observation of radio flaring from GRS 1915+105}",
      journal = {The Astronomer's Telegram},
         year = 2019,
        month = may,
       volume = {12773},
        pages = {1},
       adsurl = {https://ui.adsabs.harvard.edu/abs/2019ATel12773....1M}
}

@ARTICLE{Koljonen2019,
       author = {{Koljonen}, Karri and {Vera}, Rafael and {Lahteenmaki}, Anne and {Tornikoski}, Merja},
        title = "{A sudden radio brightening of GRS 1915+105 with Metsahovi Radio Observatory at 37 GHz}",
      journal = {The Astronomer's Telegram},
         year = 2019,
        month = jun,
       volume = {12839},
        pages = {1},
       adsurl = {https://ui.adsabs.harvard.edu/abs/2019ATel12839....1K}
}

@ARTICLE{Homan2005,
       author = {{Homan}, Jeroen and {Belloni}, Tomaso},
        title = "{The Evolution of Black Hole States}",
      journal = {\apss},
         year = 2005,
        month = nov,
       volume = {300},
       number = {1-3},
        pages = {107-117},
          doi = {10.1007/s10509-005-1197-4},
archivePrefix = {arXiv},
       eprint = {astro-ph/0412597},
 primaryClass = {astro-ph},
       adsurl = {https://ui.adsabs.harvard.edu/abs/2005Ap&SS.300..107H}
}

@ARTICLE{Nathan2022,
       author = {{Nathan}, Edward and {Ingram}, Adam and {Homan}, Jeroen and {Huppenkothen}, Daniela and {Uttley}, Phil and {van der Klis}, Michiel and {Motta}, Sara and {Altamirano}, Diego and {Middleton}, Matthew},
        title = "{Phase-resolved spectroscopy of a quasi-periodic oscillation in the black hole X-ray binary GRS 1915+105 with NICER and NuSTAR}",
      journal = {\mnras},
         year = 2022,
        month = mar,
       volume = {511},
       number = {1},
        pages = {255-279},
          doi = {10.1093/mnras/stab3803},
archivePrefix = {arXiv},
       eprint = {2201.01765},
 primaryClass = {astro-ph.HE},
       adsurl = {https://ui.adsabs.harvard.edu/abs/2022MNRAS.511..255N}
}

@ARTICLE{Belloni2011,
       author = {{Belloni}, T.~M. and {Motta}, S.~E. and {Mu{\~n}oz-Darias}, T.},
        title = "{Black hole transients}",
      journal = {Bulletin of the Astronomical Society of India},
         year = 2011,
        month = sep,
       volume = {39},
       number = {3},
        pages = {409-428},
          doi = {10.48550/arXiv.1109.3388},
archivePrefix = {arXiv},
       eprint = {1109.3388},
 primaryClass = {astro-ph.HE},
       adsurl = {https://ui.adsabs.harvard.edu/abs/2011BASI...39..409B}
}

@INCOLLECTION{Belloni2010,
       author = {{Belloni}, T.~M.},
        title = "{States and Transitions in Black Hole Binaries}",
    booktitle = {Lecture Notes in Physics, Berlin Springer Verlag},
         year = 2010,
       editor = {{Belloni}, Tomaso},
       volume = {794},
        pages = {53},
          doi = {10.1007/978-3-540-76937-8_3},
       adsurl = {https://ui.adsabs.harvard.edu/abs/2010LNP...794...53B}
}

@INCOLLECTION{vanderKlis2006,
       author = {{van der Klis}, M.},
        title = "{Rapid X-ray Variability}",
    booktitle = {Compact stellar X-ray sources},
         year = 2006,
       volume = {39},
        pages = {39-112},
       adsurl = {https://ui.adsabs.harvard.edu/abs/2006csxs.book...39V}
}

@ARTICLE{Wijnands1999,
       author = {{Wijnands}, Rudy and {Homan}, Jeroen and {van der Klis}, Michiel},
        title = "{The Complex Phase-Lag Behavior of the 3-12 HZ Quasi-Periodic Oscillations during the Very High State of XTE J1550-564}",
      journal = {\apjl},
         year = 1999,
        month = nov,
       volume = {526},
       number = {1},
        pages = {L33-L36},
          doi = {10.1086/312365},
archivePrefix = {arXiv},
       eprint = {astro-ph/9909515},
 primaryClass = {astro-ph},
       adsurl = {https://ui.adsabs.harvard.edu/abs/1999ApJ...526L..33W}
}

@ARTICLE{Remillard2002,
       author = {{Remillard}, Ronald A. and {Sobczak}, Gregory J. and {Muno}, Michael P. and {McClintock}, Jeffrey E.},
        title = "{Characterizing the Quasi-periodic Oscillation Behavior of the X-Ray Nova XTE J1550-564}",
      journal = {\apj},
         year = 2002,
        month = jan,
       volume = {564},
       number = {2},
        pages = {962-973},
          doi = {10.1086/324276},
archivePrefix = {arXiv},
       eprint = {astro-ph/0105508},
 primaryClass = {astro-ph},
       adsurl = {https://ui.adsabs.harvard.edu/abs/2002ApJ...564..962R}
}

@ARTICLE{Remillard2006,
       author = {{Remillard}, Ronald A. and {McClintock}, Jeffrey E.},
        title = "{X-Ray Properties of Black-Hole Binaries}",
      journal = {\araa},
         year = 2006,
        month = sep,
       volume = {44},
       number = {1},
        pages = {49-92},
          doi = {10.1146/annurev.astro.44.051905.092532},
archivePrefix = {arXiv},
       eprint = {astro-ph/0606352},
 primaryClass = {astro-ph},
       adsurl = {https://ui.adsabs.harvard.edu/abs/2006ARA&A..44...49R}
}

@ARTICLE{Wilkinson2009,
       author = {{Wilkinson}, Tony and {Uttley}, Philip},
        title = "{Accretion disc variability in the hard state of black hole X-ray binaries}",
      journal = {\mnras},
         year = 2009,
        month = aug,
       volume = {397},
       number = {2},
        pages = {666-676},
          doi = {10.1111/j.1365-2966.2009.15008.x},
archivePrefix = {arXiv},
       eprint = {0905.0587},
 primaryClass = {astro-ph.HE},
       adsurl = {https://ui.adsabs.harvard.edu/abs/2009MNRAS.397..666W}
}

@ARTICLE{Zhou2022,
       author = {{Zhou}, Deng-Ke and {Zhang}, Shuang-Nan and {Song}, Li-Ming and {Qu}, Jin-Lu and {Zhang}, Liang and {Ma}, Xiang and {Tuo}, You-Li and {Ge}, Ming-Yu and {Wang}, Yanan and {Zhang}, Shu and {Tao}, Lian},
        title = "{Determination of QPO properties in the presence of strong broad-band noise: a case study on the data of MAXI J1820+070}",
      journal = {\mnras},
         year = 2022,
        month = sep,
       volume = {515},
       number = {2},
        pages = {1914-1926},
          doi = {10.1093/mnras/stac1789},
archivePrefix = {arXiv},
       eprint = {2206.12905},
 primaryClass = {astro-ph.HE},
       adsurl = {https://ui.adsabs.harvard.edu/abs/2022MNRAS.515.1914Z}
}

@ARTICLE{Konig2024,
       author = {{K{\"o}nig}, Ole and {Mastroserio}, Guglielmo and {Dauser}, Thomas and {M{\'e}ndez}, Mariano and {Wang}, Jingyi and {Garc{\'\i}a}, Javier A. and {Steiner}, James F. and {Pottschmidt}, Katja and {Ballhausen}, Ralf and {Connors}, Riley M. and {Garc{\'\i}a}, Federico and {Grinberg}, Victoria and {Horn}, David and {Ingram}, Adam and {Kara}, Erin and {Kallman}, Timothy R. and {Lucchini}, Matteo and {Nathan}, Edward and {Nowak}, Michael A. and {Thalhammer}, Philipp and {van der Klis}, Michiel and {Wilms}, J{\"o}rn},
        title = "{Long term variability of Cygnus X-1. VIII. A spectral-timing look at low energies with NICER}",
      journal = {\aap},
         year = 2024,
        month = jul,
       volume = {687},
          eid = {A284},
        pages = {A284},
          doi = {10.1051/0004-6361/202449333},
archivePrefix = {arXiv},
       eprint = {2405.07754},
 primaryClass = {astro-ph.HE},
       adsurl = {https://ui.adsabs.harvard.edu/abs/2024A&A...687A.284K}
}

@ARTICLE{Fogantini2025,
       author = {{Fogantini}, Federico A. and {Garc{\'\i}a}, Federico and {M{\'e}ndez}, Mariano and {K{\"o}nig}, Ole and {Wilms}, Joern},
        title = "{A hidden quasi-periodic oscillation in Cygnus X-1 revealed by NICER}",
      journal = {\aap},
         year = 2025,
        month = apr,
       volume = {696},
          eid = {A237},
        pages = {A237},
          doi = {10.1051/0004-6361/202453523},
archivePrefix = {arXiv},
       eprint = {2503.03078},
 primaryClass = {astro-ph.HE},
       adsurl = {https://ui.adsabs.harvard.edu/abs/2025A&A...696A.237F}
}

@INPROCEEDINGS{Zhang1993,
       author = {{Zhang}, Weiping and {Giles}, Alan B. and {Jahoda}, K. and {Soong}, Yang and {Swank}, Jean H. and {Morgan}, Edward H.},
        title = "{Laboratory performance of the proportional counter array experiment for the X-ray Timing Explorer}",
    booktitle = {EUV, X-Ray, and Gamma-Ray Instrumentation for Astronomy IV},
         year = 1993,
       editor = {{Siegmund}, Oswald H.},
       series = {Society of Photo-Optical Instrumentation Engineers (SPIE) Conference Series},
       volume = {2006},
        month = nov,
        pages = {324-333},
          doi = {10.1117/12.162845},
       adsurl = {https://ui.adsabs.harvard.edu/abs/1993SPIE.2006..324Z}
}

@ARTICLE{SunyaevTitarchuk1980,
       author = {{Sunyaev}, R.~A. and {Titarchuk}, L.~G.},
        title = "{Comptonization of X-Rays in Plasma Clouds - Typical Radiation Spectra}",
      journal = {\aap},
         year = 1980,
        month = jun,
       volume = {86},
        pages = {121},
       adsurl = {https://ui.adsabs.harvard.edu/abs/1980A&A....86..121S}
}

@ARTICLE{ShakuraSunyaev1973,
       author = {{Shakura}, N.~I. and {Sunyaev}, R.~A.},
        title = "{Black holes in binary systems. Observational appearance.}",
      journal = {\aap},
         year = 1973,
        month = jan,
       volume = {24},
        pages = {337-355},
       adsurl = {https://ui.adsabs.harvard.edu/abs/1973A&A....24..337S}
}

@ARTICLE{Karpouzas2021,
       author = {{Karpouzas}, Konstantinos and {M{\'e}ndez}, Mariano and {Garc{\'\i}a}, Federico and {Zhang}, Liang and {Altamirano}, Diego and {Belloni}, Tomaso and {Zhang}, Yuexin},
        title = "{A variable corona for GRS 1915+105}",
      journal = {\mnras},
         year = 2021,
        month = jun,
       volume = {503},
       number = {4},
        pages = {5522-5533},
          doi = {10.1093/mnras/stab827},
archivePrefix = {arXiv},
       eprint = {2103.09675},
 primaryClass = {astro-ph.HE},
       adsurl = {https://ui.adsabs.harvard.edu/abs/2021MNRAS.503.5522K}
}

@ARTICLE{Ingram2009,
       author = {{Ingram}, Adam and {Done}, Chris and {Fragile}, P. Chris},
        title = "{Low-frequency quasi-periodic oscillations spectra and Lense-Thirring precession}",
      journal = {\mnras},
         year = 2009,
        month = jul,
       volume = {397},
       number = {1},
        pages = {L101-L105},
          doi = {10.1111/j.1745-3933.2009.00693.x},
archivePrefix = {arXiv},
       eprint = {0901.1238},
 primaryClass = {astro-ph.SR},
       adsurl = {https://ui.adsabs.harvard.edu/abs/2009MNRAS.397L.101I}
}

@ARTICLE{Kara2019,
       author = {{Kara}, E. and {Steiner}, J.~F. and {Fabian}, A.~C. and {Cackett}, E.~M. and {Uttley}, P. and {Remillard}, R.~A. and {Gendreau}, K.~C. and {Arzoumanian}, Z. and {Altamirano}, D. and {Eikenberry}, S. and {Enoto}, T. and {Homan}, J. and {Neilsen}, J. and {Stevens}, A.~L.},
        title = "{The corona contracts in a black-hole transient}",
      journal = {\nat},
         year = 2019,
        month = jan,
       volume = {565},
       number = {7738},
        pages = {198-201},
          doi = {10.1038/s41586-018-0803-x},
archivePrefix = {arXiv},
       eprint = {1901.03877},
 primaryClass = {astro-ph.HE},
       adsurl = {https://ui.adsabs.harvard.edu/abs/2019Natur.565..198K}
}

@ARTICLE{Mastroserio2019,
       author = {{Mastroserio}, Guglielmo and {Ingram}, Adam and {van der Klis}, Michiel},
        title = "{An X-ray reverberation mass measurement of Cygnus X-1}",
      journal = {\mnras},
         year = 2019,
        month = sep,
       volume = {488},
       number = {1},
        pages = {348-361},
          doi = {10.1093/mnras/stz1727},
archivePrefix = {arXiv},
       eprint = {1906.08266},
 primaryClass = {astro-ph.HE},
       adsurl = {https://ui.adsabs.harvard.edu/abs/2019MNRAS.488..348M}
}

@ARTICLE{Belloni2002,
       author = {{Belloni}, Tomaso and {Psaltis}, Dimitrios and {van der Klis}, Michiel},
        title = "{A Unified Description of the Timing Features of Accreting X-Ray Binaries}",
      journal = {\apj},
         year = 2002,
        month = jun,
       volume = {572},
       number = {1},
        pages = {392-406},
          doi = {10.1086/340290},
archivePrefix = {arXiv},
       eprint = {astro-ph/0202213},
 primaryClass = {astro-ph},
       adsurl = {https://ui.adsabs.harvard.edu/abs/2002ApJ...572..392B}
}

@ARTICLE{Motta2012,
       author = {{Motta}, S. and {Homan}, J. and {Mu{\~n}oz Darias}, T. and {Casella}, P. and {Belloni}, T.~M. and {Hiemstra}, B. and {M{\'e}ndez}, M.},
        title = "{Discovery of two simultaneous non-harmonically related quasi-periodic oscillations in the 2005 outburst of the black hole binary GRO J1655-40}",
      journal = {\mnras},
         year = 2012,
        month = nov,
       volume = {427},
       number = {1},
        pages = {595-606},
          doi = {10.1111/j.1365-2966.2012.22037.x},
archivePrefix = {arXiv},
       eprint = {1209.0327},
 primaryClass = {astro-ph.HE},
       adsurl = {https://ui.adsabs.harvard.edu/abs/2012MNRAS.427..595M}
}

@ARTICLE{Belloni2014,
       author = {{Belloni}, Tomaso M. and {Stella}, Luigi},
        title = "{Fast Variability from Black-Hole Binaries}",
      journal = {\ssr},
         year = 2014,
        month = sep,
       volume = {183},
       number = {1-4},
        pages = {43-60},
          doi = {10.1007/s11214-014-0076-0},
archivePrefix = {arXiv},
       eprint = {1407.7373},
 primaryClass = {astro-ph.HE},
       adsurl = {https://ui.adsabs.harvard.edu/abs/2014SSRv..183...43B}
}

@article{Ingram2013,
    author = {Ingram, Adam and Klis, Michiel van der},
    title = "{An exact analytic treatment of propagating mass accretion rate fluctuations in X-ray binaries}",
    journal = {Monthly Notices of the Royal Astronomical Society},
    volume = {434},
    number = {2},
    pages = {1476-1485},
    year = {2013},
    month = {07},
    issn = {0035-8711},
    doi = {10.1093/mnras/stt1107},
    url = {https://doi.org/10.1093/mnras/stt1107},
    eprint = {https://academic.oup.com/mnras/article-pdf/434/2/1476/18497264/stt1107.pdf},
}

@ARTICLE{Bellavita2022,
       author = {{Bellavita}, Candela and {Garc{\'\i}a}, Federico and {M{\'e}ndez}, Mariano and {Karpouzas}, Konstantinos},
        title = "{vKompth: a variable Comptonization model for low-frequency quasi-periodic oscillations in black hole X-ray binaries}",
      journal = {\mnras},
         year = 2022,
        month = sep,
       volume = {515},
       number = {2},
        pages = {2099-2109},
          doi = {10.1093/mnras/stac1922},
archivePrefix = {arXiv},
       eprint = {2206.13609},
 primaryClass = {astro-ph.HE},
       adsurl = {https://ui.adsabs.harvard.edu/abs/2022MNRAS.515.2099B}
}

@ARTICLE{Karpouzas2020,
       author = {{Karpouzas}, Konstantinos and {M{\'e}ndez}, Mariano and {Ribeiro}, Evandro M. and {Altamirano}, Diego and {Blaes}, Omer and {Garc{\'\i}a}, Federico},
        title = "{The Comptonizing medium of the neutron star in 4U 1636 - 53 through its lower kilohertz quasi-periodic oscillations}",
      journal = {\mnras},
         year = 2020,
        month = feb,
       volume = {492},
       number = {1},
        pages = {1399-1415},
          doi = {10.1093/mnras/stz3502},
archivePrefix = {arXiv},
       eprint = {1912.05380},
 primaryClass = {astro-ph.HE},
       adsurl = {https://ui.adsabs.harvard.edu/abs/2020MNRAS.492.1399K}
}

@ARTICLE{Fender2004,
       author = {{Fender}, R.~P. and {Belloni}, T.~M. and {Gallo}, E.},
        title = "{Towards a unified model for black hole X-ray binary jets}",
      journal = {\mnras},
         year = 2004,
        month = dec,
       volume = {355},
       number = {4},
        pages = {1105-1118},
          doi = {10.1111/j.1365-2966.2004.08384.x},
archivePrefix = {arXiv},
       eprint = {astro-ph/0409360},
 primaryClass = {astro-ph},
       adsurl = {https://ui.adsabs.harvard.edu/abs/2004MNRAS.355.1105F}
}

@ARTICLE{belloni1990,
       author = {{Belloni}, T. and {Hasinger}, G.},
        title = "{An atlas of aperiodic variability in HMXB.}",
      journal = {\aap},
         year = 1990,
        month = apr,
       volume = {230},
        pages = {103-119},
       adsurl = {https://ui.adsabs.harvard.edu/abs/1990A&A...230..103B}
}

@ARTICLE{leahy1983,
       author = {{Leahy}, D.~A. and {Darbro}, W. and {Elsner}, R.~F. and {Weisskopf}, M.~C. and {Sutherland}, P.~G. and {Kahn}, S. and {Grindlay}, J.~E.},
        title = "{On searches for pulsed emission with application to four globular cluster X-ray sources : NGC 1851, 6441, 6624 and 6712.}",
      journal = {\apj},
         year = 1983,
        month = mar,
       volume = {266},
        pages = {160-170},
          doi = {10.1086/160766},
       adsurl = {https://ui.adsabs.harvard.edu/abs/1983ApJ...266..160L}
}

@ARTICLE{Arevalo2006,
       author = {{Ar{\'e}valo}, P. and {Uttley}, P.},
        title = "{Investigating a fluctuating-accretion model for the spectral-timing properties of accreting black hole systems}",
      journal = {\mnras},
         year = 2006,
        month = apr,
       volume = {367},
       number = {2},
        pages = {801-814},
          doi = {10.1111/j.1365-2966.2006.09989.x},
archivePrefix = {arXiv},
       eprint = {astro-ph/0512394},
 primaryClass = {astro-ph},
       adsurl = {https://ui.adsabs.harvard.edu/abs/2006MNRAS.367..801A}
}

@ARTICLE{motta2021,
       author = {{Motta}, S.~E. and {Kajava}, J.~J.~E. and {Giustini}, M. and {Williams}, D.~R.~A. and {Del Santo}, M. and {Fender}, R. and {Green}, D.~A. and {Heywood}, I. and {Rhodes}, L. and {Segreto}, A. and {Sivakoff}, G. and {Woudt}, P.~A.},
        title = "{Observations of a radio-bright, X-ray obscured GRS 1915+105}",
      journal = {\mnras},
         year = 2021,
        month = may,
       volume = {503},
       number = {1},
        pages = {152-161},
          doi = {10.1093/mnras/stab511},
archivePrefix = {arXiv},
       eprint = {2101.01187},
 primaryClass = {astro-ph.HE},
       adsurl = {https://ui.adsabs.harvard.edu/abs/2021MNRAS.503..152M}
}

@ARTICLE{Miller2020,
       author = {{Miller}, J.~M. and {Zoghbi}, A. and {Raymond}, J. and {Balakrishnan}, M. and {Brenneman}, L. and {Cackett}, E. and {Draghis}, P. and {Fabian}, A.~C. and {Gallo}, E. and {Kaastra}, J. and {Kallman}, T. and {Kammoun}, E. and {Motta}, S.~E. and {Proga}, D. and {Reynolds}, M.~T. and {Trueba}, N.},
        title = "{An Obscured, Seyfert 2-like State of the Stellar-mass Black Hole GRS 1915+105 Caused by Failed Disk Winds}",
      journal = {\apj},
         year = 2020,
        month = nov,
       volume = {904},
       number = {1},
          eid = {30},
        pages = {30},
          doi = {10.3847/1538-4357/abbb31},
archivePrefix = {arXiv},
       eprint = {2007.07005},
 primaryClass = {astro-ph.HE},
       adsurl = {https://ui.adsabs.harvard.edu/abs/2020ApJ...904...30M}
}

@ARTICLE{eijnden2017,
       author = {{van den Eijnden}, J. and {Ingram}, A. and {Uttley}, P. and {Motta}, S.~E. and {Belloni}, T.~M. and {Gardenier}, D.~W.},
        title = "{Inclination dependence of QPO phase lags in black hole X-ray binaries}",
      journal = {\mnras},
         year = 2017,
        month = jan,
       volume = {464},
       number = {3},
        pages = {2643-2659},
          doi = {10.1093/mnras/stw2634},
archivePrefix = {arXiv},
       eprint = {1610.03469},
 primaryClass = {astro-ph.HE},
       adsurl = {https://ui.adsabs.harvard.edu/abs/2017MNRAS.464.2643V}
}

@ARTICLE{Reig2000,
       author = {{Reig}, P. and {Belloni}, T. and {van der Klis}, M. and {M{\'e}ndez}, M. and {Kylafis}, N.~D. and {Ford}, E.~C.},
        title = "{Phase Lag Variability Associated with the 0.5-10 HZ Quasi-Periodic Oscillations in GRS 1915+105}",
      journal = {\apj},
         year = 2000,
        month = oct,
       volume = {541},
       number = {2},
        pages = {883-888},
          doi = {10.1086/309469},
       adsurl = {https://ui.adsabs.harvard.edu/abs/2000ApJ...541..883R}
}

@ARTICLE{Pahari2013,
       author = {{Pahari}, Mayukh and {Neilsen}, Joseph and {Yadav}, J.~S. and {Misra}, Ranjeev and {Uttley}, Phil},
        title = "{Comparison of Time/Phase Lags in the Hard State and Plateau State of GRS 1915+105}",
      journal = {\apj},
         year = 2013,
        month = dec,
       volume = {778},
       number = {2},
          eid = {136},
        pages = {136},
          doi = {10.1088/0004-637X/778/2/136},
archivePrefix = {arXiv},
       eprint = {1310.3037},
 primaryClass = {astro-ph.HE},
       adsurl = {https://ui.adsabs.harvard.edu/abs/2013ApJ...778..136P}
}

@ARTICLE{Qu2010,
       author = {{Qu}, J.~L. and {Lu}, F.~J. and {Lu}, Y. and {Song}, L.~M. and {Zhang}, S. and {Ding}, G.~Q. and {Wang}, J.~M.},
        title = "{The Energy Dependence of the Centroid Frequency and Phase Lag of the Quasi-periodic Oscillations in GRS 1915+105}",
      journal = {\apj},
         year = 2010,
        month = feb,
       volume = {710},
       number = {1},
        pages = {836-842},
          doi = {10.1088/0004-637X/710/1/836},
archivePrefix = {arXiv},
       eprint = {0912.4769},
 primaryClass = {astro-ph.HE},
       adsurl = {https://ui.adsabs.harvard.edu/abs/2010ApJ...710..836Q}
}

\appendix
\section{Properties of the selected \NICER dataset}
\label{sec:appendix}

In Table~\ref{tab:tabla}, we show the full list of \NICER observations of \grs used in this work. For each observation we report the starting time, number of selected 65-s segments after filtering, count rate in the $0.3-12$~keV, and hardness ratio using the $5-12$~keV and $2-5$~keV energy bands. We also include the properties of the detected QPOs: centroid frequency, FWHM, fractional rms amplitude in the $2-5$~keV and $5-12$~keV bands, and broadband phase lag. Finally, we report the photon index $\Gamma$ from the spectral analysis.

\begin{table*}
\centering
 \caption{Summary of the \grs observations analysed in this work, including the properties of the detected type-C QPOs.}
 \resizebox{\textwidth}{!}{
 \begin{tabular}{ccccccccccc}
  \hline
  \noalign{\smallskip}
  Observation & Start time & 65-sec & Count rate$^a$ & ${\rm HR}^b$ & $\nu_{\rm QPO}$ & FWHM & ${\rm QPO~rms_{~2-5~keV}}$ & ${\rm QPO~rms_{~5-12~keV}}$ & ${\rm Broadband~ phase~ lag}^c$ &$\Gamma$\\
  & MJD$-$58420 & segments & (counts/s) & & (Hz) & (Hz) & (\%) & (\%) & (rad) &  \\
  \noalign{\smallskip}
  \hline
1103010138 & 38.998 & 28 & 607.6 & 0.35 & $3.865 \pm 0.020$ & $0.76 \pm 0.07$ & $6.3 \pm 0.2$ & $13.0 \pm 0.5$ & $-0.11 \pm 0.06$  & $2.46^{+0.03}_{-0.05}$ \\
1103010139 & 43.884 & 21 & 558.1 & 0.36 & $3.562 \pm 0.016$ & $0.56 \pm 0.06$ & $7.3 \pm 0.3$ & $12.2 \pm 0.6$ & $-0.21 \pm 0.06$  & $2.32^{+0.02}_{-0.03}$ \\
1103010140 & 44.015 & 23 & 541.6 & 0.37 & $3.382 \pm 0.016$ & $0.50 \pm 0.06$ & $7.2 \pm 0.3$ & $11.5 \pm 0.6$ & $-0.15 \pm 0.07$  & $2.27^{+0.01}_{-0.02}$ \\
1103010142 & 60.123 & 16 & 447.4 & 0.37 & $3.085 \pm 0.016$ & $0.41 \pm 0.06$ & $7.2 \pm 0.3$ & $11.7 \pm 0.6$ & $-0.20 \pm 0.08$  & $2.21 \pm 0.02$ \\
1103010143 & 61.847 & 17 & 458.9 & 0.36 & $3.595 \pm 0.018$ & $0.55 \pm 0.07$ & $7.6 \pm 0.3$ & $12.0 \pm 0.7$ & $-0.27 \pm 0.08$  & $2.32^{+0.03}_{-0.04}$ \\
1103010144 & 62.504 & 10 & 432.3 & 0.37 & $3.198 \pm 0.021$ & $0.32 \pm 0.05$ & $6.1 \pm 0.4$ & $11.5 \pm 0.9$ & $-0.34 \pm 0.11$  & $2.19^{+0.01}_{-0.03}$ \\
1103010145 & 63.330 & 17 & 413.4 & 0.37 & $3.043 \pm 0.017$ & $0.46 \pm 0.07$ & $7.5 \pm 0.3$ & $12.6 \pm 0.7$ & $-0.14 \pm 0.08$  & $2.16^{+0.01}_{-0.02}$ \\
1103010147 & 65.962 & 13 & 384.9 & 0.38 & $2.736 \pm 0.015$ & $0.30 \pm 0.04$ & $7.0 \pm 0.4$ & $10.8 \pm 0.8$ & $-0.14 \pm 0.10$  & $2.10 \pm 0.01$ \\
1103010148 & 66.669 & 12 & 368.5 & 0.39 & $2.472 \pm 0.011$ & $0.34 \pm 0.04$ & $8.8 \pm 0.4$ & $15.1 \pm 0.8$ & $-0.05 \pm 0.08$  & $2.01^{+0.02}_{-0.03}$ \\
1103010149 & 68.664 & 15 & 339.0 & 0.42 & $2.192 \pm 0.010$ & $0.18 \pm 0.03$ & $6.9 \pm 0.4$ & $10.8 \pm 0.8$ & $-0.24 \pm 0.10$  & $1.98 \pm 0.02$ \\
1103010150 & 69.500 & 24 & 337.1 & 0.42 & $2.211 \pm 0.008$ & $0.27 \pm 0.03$ & $8.0 \pm 0.3$ & $12.3 \pm 0.6$ & $-0.10 \pm 0.07$  & $1.99 \pm 0.02$ \\
1103010152 & 71.366 & 27 & 377.9 & 0.38 & $3.106 \pm 0.015$ & $0.44 \pm 0.04$ & $6.9 \pm 0.3$ & $13.4 \pm 0.6$ & $-0.14 \pm 0.07$  & $2.17^{+0.01}_{-0.02}$ \\
1103010153 & 72.468 & 21 & 356.7 & 0.38 & $2.887 \pm 0.010$ & $0.24 \pm 0.04$ & $6.1 \pm 0.3$ & $11.1 \pm 0.7$ & $-0.10 \pm 0.09$  & $2.12^{+0.03}_{-0.01}$ \\
1103010154 & 73.432 & 22 & 351.6 & 0.38 & $2.942 \pm 0.012$ & $0.26 \pm 0.04$ & $6.1 \pm 0.3$ & $10.7 \pm 0.7$ & $-0.17 \pm 0.09$  & $2.13^{+0.02}_{-0.01}$ \\
1103010155 & 74.583 & 14 & 350.3 & 0.38 & $2.793 \pm 0.013$ & $0.29 \pm 0.04$ & $7.2 \pm 0.4$ & $11.7 \pm 0.8$ & $-0.18 \pm 0.10$  & $2.11 \pm 0.02$ \\
1103010156 & 75.484 & 34 & 336.1 & 0.39 & $2.638 \pm 0.010$ & $0.38 \pm 0.03$ & $8.2 \pm 0.3$ & $13.6 \pm 0.5$ & $-0.11 \pm 0.06$  & $2.08 \pm 0.02$ \\
1103010157 & 77.491 & 248 & 301.6 & 0.41 & $2.160 \pm 0.003$ & $0.27 \pm 0.01$ & $8.8 \pm 0.1$ & $12.4 \pm 0.2$ & $-0.13 \pm 0.02$  & $1.99 \pm 0.01$ \\
1103010158 & 77.994 & 77 & 302.6 & 0.42 & $2.176 \pm 0.006$ & $0.32 \pm 0.02$ & $9.1 \pm 0.2$ & $13.0 \pm 0.4$ & $-0.09 \pm 0.04$  & $1.98 \pm 0.01$ \\
1103010159 & 99.045 & 21 & 226.2 & 0.43 & $1.918 \pm 0.010$ & $0.26 \pm 0.03$ & $9.3 \pm 0.4$ & $13.2 \pm 0.7$ & $-0.10 \pm 0.08$  & $1.94 \pm 0.02$ \\
1103010160 & 108.830 & 10 & 202.6 & 0.43 & $1.947 \pm 0.014$ & $0.22 \pm 0.03$ & $8.0 \pm 0.6$ & $12.0 \pm 1.1$ & $\phantom{-}0.07 \pm 0.12$  & $1.92 \pm 0.03$ \\
1103010161 & 109.147 & 43 & 193.9 & 0.43 & $1.856 \pm 0.006$ & $0.15 \pm 0.02$ & $7.2 \pm 0.3$ & $11.0 \pm 0.6$ & $-0.06 \pm 0.07$  & $1.91 \pm 0.02$ \\
1103010162 & 110.839 & 24 & 193.8 & 0.43 & $1.890 \pm 0.007$ & $0.16 \pm 0.02$ & $7.4 \pm 0.4$ & $10.0 \pm 0.8$ & $-0.09 \pm 0.11$  & $1.82 \pm 0.01$ \\
1103010163 & 112.125 & 33 & 198.9 & 0.42 & $2.143 \pm 0.011$ & $0.25 \pm 0.03$ & $7.4 \pm 0.4$ & $11.3 \pm 0.7$ & $-0.11 \pm 0.09$  & $1.96 \pm 0.02$ \\
1103010164 & 113.539 & 30 & 183.2 & 0.44 & $1.761 \pm 0.008$ & $0.18 \pm 0.02$ & $8.5 \pm 0.4$ & $12.2 \pm 0.7$ & $\phantom{-}0.04 \pm 0.08$  & $1.88 \pm 0.02$ \\
1103010165 & 114.695 & 17 & 191.0 & 0.43 & $1.981 \pm 0.014$ & $0.14 \pm 0.03$ & $7.1 \pm 0.5$ & $9.4 \pm 1.0$ & $-0.33 \pm 0.14$  & $1.90 \pm 0.02$ \\
1103010166 & 115.595 & 21 & 190.2 & 0.43 & $2.050 \pm 0.005$ & $0.11 \pm 0.10$ & $6.2 \pm 0.4$ & $10.7 \pm 0.8$ & $\phantom{-}0.03 \pm 0.12$  & $1.93 \pm 0.02$ \\
1103010168 & 126.986 & 18 & 151.6 & 0.46 & $1.484 \pm 0.007$ & $0.14 \pm 0.02$ & $8.9 \pm 0.5$ & $12.0 \pm 0.9$ & $-0.06 \pm 0.10$  & $1.87 \pm 0.02$ \\
1103010169 & 127.051 & 18 & 154.4 & 0.46 & $1.494 \pm 0.007$ & $0.17 \pm 0.03$ & $9.6 \pm 0.5$ & $14.2 \pm 0.9$ & $-0.01 \pm 0.09$  & $1.87 \pm 0.02$ \\
1103010170 & 128.016 & 36 & 160.9 & 0.45 & $1.741 \pm 0.006$ & $0.12 \pm 0.02$ & $6.5 \pm 0.4$ & $9.8 \pm 0.7$ & $\phantom{-}0.06 \pm 0.11$  & $1.87 \pm 0.02$ \\
1103010171 & 129.046 & 38 & 164.1 & 0.44 & $1.853 \pm 0.005$ & $0.12 \pm 0.01$ & $7.6 \pm 0.3$ & $10.3 \pm 0.6$ & $-0.06 \pm 0.08$  & $1.88 \pm 0.02$ \\
1103010172 & 130.011 & 60 & 165.9 & 0.43 & $2.000 \pm 0.009$ & $0.31 \pm 0.03$ & $9.8 \pm 0.3$ & $12.9 \pm 0.5$ & $-0.01 \pm 0.06$  & $1.89 \pm 0.02$ \\
1103010174 & 132.070 & 13 & 157.6 & 0.44 & $1.823 \pm 0.011$ & $0.18 \pm 0.03$ & $8.8 \pm 0.6$ & $11.0 \pm 1.1$ & $-0.08 \pm 0.13$  & $1.89 \pm 0.03$ \\
1103010179 & 141.846 & 21 & 151.1 & 0.47 & $1.564 \pm 0.011$ & $0.21 \pm 0.03$ & $9.2 \pm 0.5$ & $11.7 \pm 0.9$ & $\phantom{-}0.17 \pm 0.10$  & $1.77 \pm 0.02$ \\
1103010180 & 143.133 & 18 & 145.2 & 0.46 & $1.547 \pm 0.009$ & $0.11 \pm 0.03$ & $7.4 \pm 0.5$ & $11.9 \pm 1.0$ & $\phantom{-}0.04 \pm 0.12$  & $1.82 \pm 0.02$ \\
1103010181 & 147.188 & 36 & 152.5 & 0.45 & $1.798 \pm 0.016$ & $0.44 \pm 0.06$ & $10.8 \pm 0.4$ & $12.0 \pm 0.8$ & $-0.15 \pm 0.08$  & $1.84 \pm 0.02$ \\
1103010184 & 154.484 & 13 & 145.6 & 0.45 & $1.728 \pm 0.012$ & $0.15 \pm 0.04$ & $7.6 \pm 0.6$ & $10.6 \pm 1.1$ & $\phantom{-}0.14 \pm 0.15$  & $1.84 \pm 0.03$ \\
1103010186 & 158.274 & 26 & 149.9 & 0.45 & $1.867 \pm 0.006$ & $0.11 \pm 0.02$ & $7.7 \pm 0.4$ & $11.0 \pm 0.7$ & $\phantom{-}0.12 \pm 0.10$  & $1.81^{+0.02}_{-0.03}$ \\
1103010187 & 159.368 & 11 & 146.8 & 0.45 & $1.770 \pm 0.008$ & $0.04 \pm 0.04$ & $6.3 \pm 0.8$ & $6.6 \pm 1.3$ & $\phantom{-}0.22 \pm 0.24$  & $1.84 \pm 0.03$ \\
1103010189 & 161.304 & 19 & 141.8 & 0.44 & $1.791 \pm 0.008$ & $0.14 \pm 0.02$ & $8.7 \pm 0.6$ & $10.5 \pm 1.1$ & $\phantom{-}0.07 \pm 0.14$  & $1.87 \pm 0.02$ \\
1103010190 & 162.268 & 18 & 150.2 & 0.43 & $2.018 \pm 0.008$ & $0.13 \pm 0.02$ & $7.6 \pm 0.5$ & $12.0 \pm 0.9$ & $\phantom{-}0.03 \pm 0.11$  & $1.92 \pm 0.02$ \\
1103010192 & 165.352 & 29 & 150.5 & 0.44 & $2.024 \pm 0.007$ & $0.18 \pm 0.02$ & $9.9 \pm 0.4$ & $11.9 \pm 0.7$ & $-0.12 \pm 0.08$  & $1.84 \pm 0.01$ \\
1103010193 & 178.219 & 24 & 138.4 & 0.44 & $1.829 \pm 0.007$ & $0.16 \pm 0.02$ & $9.4 \pm 0.5$ & $10.7 \pm 0.8$ & $-0.15 \pm 0.10$  & $1.88 \pm 0.02$ \\
1103010194 & 211.691 & 20 & 127.7 & 0.42 & $2.155 \pm 0.006$ & $0.04 \pm 0.02$ & $5.1 \pm 0.5$ & $5.8 \pm 1.1$ & $\phantom{-}0.08 \pm 0.23$  & $1.96 \pm 0.03$ \\
1103010195 & 215.808 & 38 & 119.2 & 0.46 & $1.655 \pm 0.005$ & $0.17 \pm 0.03$ & $9.5 \pm 0.4$ & $12.4 \pm 0.7$ & $-0.07 \pm 0.08$  & $1.90 \pm 0.03$ \\
1103010196 & 216.195 & 37 & 115.5 & 0.46 & $1.612 \pm 0.004$ & $0.10 \pm 0.02$ & $7.8 \pm 0.4$ & $9.9 \pm 0.8$ & $\phantom{-}0.03 \pm 0.11$  & $1.83^{+0.02}_{-0.06}$ \\
1103010197 & 220.314 & 78 & 108.5 & 0.48 & $1.253 \pm 0.004$ & $0.10 \pm 0.01$ & $7.8 \pm 0.3$ & $9.1 \pm 0.6$ & $-0.08 \pm 0.08$  & $1.79 \pm 0.02$ \\
1103010198 & 224.756 & 92 & 117.2 & 0.46 & $1.507 \pm 0.005$ & $0.19 \pm 0.02$ & $9.0 \pm 0.3$ & $11.5 \pm 0.5$ & $-0.07 \pm 0.06$  & $1.83 \pm 0.02$ \\
1103010199 & 225.013 & 126 & 123.3 & 0.46 & $1.628 \pm 0.004$ & $0.21 \pm 0.01$ & $9.5 \pm 0.2$ & $11.7 \pm 0.4$ & $-0.07 \pm 0.05$  & $1.84 \pm 0.02$ \\
1103010201 & 226.946 & 27 & 123.2 & 0.46 & $1.611 \pm 0.008$ & $0.16 \pm 0.03$ & $8.7 \pm 0.5$ & $9.1 \pm 0.9$ & $-0.07 \pm 0.12$  & $1.71^{+0.02}_{-0.01}$ \\
1103010206 & 238.093 & 29 & 118.8 & 0.46 & $1.570 \pm 0.012$ & $0.27 \pm 0.04$ & $10.0 \pm 0.5$ & $12.5 \pm 0.9$ & $\phantom{-}0.11 \pm 0.10$  & $1.86 \pm 0.02$ \\
1103010207 & 239.189 & 24 & 119.9 & 0.46 & $1.601 \pm 0.008$ & $0.18 \pm 0.03$ & $9.6 \pm 0.5$ & $10.3 \pm 0.9$ & $-0.05 \pm 0.11$  & $1.83 \pm 0.02$ \\
1103010208 & 240.018 & 34 & 114.4 & 0.47 & $1.402 \pm 0.006$ & $0.10 \pm 0.06$ & $7.4 \pm 0.5$ & $9.0 \pm 0.8$ & $-0.25 \pm 0.12$  & $1.78 \pm 0.02$ \\
1103010209 & 249.090 & 46 & 100.5 & 0.47 & $1.409 \pm 0.010$ & $0.28 \pm 0.11$ & $8.6 \pm 0.5$ & $10.9 \pm 0.9$ & $-0.17 \pm 0.11$  & $1.82 \pm 0.02$ \\
1103010210 & 250.956 & 13 & 104.6 & 0.48 & $1.283 \pm 0.013$ & $0.08 \pm 0.03$ & $7.4 \pm 0.7$ & $9.6 \pm 1.3$ & $\phantom{-}0.23 \pm 0.18$  & $1.78 \pm 0.03$ \\
1103010211 & 251.793 & 26 & 111.5 & 0.47 & $1.573 \pm 0.007$ & $0.12 \pm 0.02$ & $8.4 \pm 0.5$ & $10.8 \pm 0.9$ & $-0.27 \pm 0.11$  & $1.85 \pm 0.02$ \\
1103010212 & 252.886 & 14 & 107.3 & 0.46 & $1.560 \pm 0.020$ & $0.22 \pm 0.05$ & $8.6 \pm 0.7$ & $10.7 \pm 1.3$ & $-0.03 \pm 0.15$  & $1.87 \pm 0.03$ \\
1103010214 & 255.203 & 14 & 104.1 & 0.47 & $1.359 \pm 0.013$ & $0.20 \pm 0.04$ & $10.9 \pm 0.7$ & $9.9 \pm 1.2$ & $\phantom{-}0.09 \pm 0.14$  & $1.82 \pm 0.03$ \\
1103010215 & 256.234 & 45 & 106.1 & 0.48 & $1.367 \pm 0.007$ & $0.19 \pm 0.02$ & $9.5 \pm 0.4$ & $10.3 \pm 0.7$ & $\phantom{-}0.03 \pm 0.08$  & $1.79 \pm 0.02$ \\
1103010216 & 257.263 & 21 & 108.3 & 0.48 & $1.358 \pm 0.005$ & $0.05 \pm 0.02$ & $6.3 \pm 0.6$ & $7.8 \pm 1.0$ & $-0.21 \pm 0.18$  & $1.74 \pm 0.03$ \\
1103010217 & 258.228 & 19 & 107.5 & 0.48 & $1.338 \pm 0.007$ & $0.07 \pm 0.02$ & $6.1 \pm 0.6$ & $8.1 \pm 1.1$ & $\phantom{-}0.04 \pm 0.19$  & $1.79 \pm 0.03$ \\
1103010219 & 260.223 & 16 & 106.4 & 0.48 & $1.404 \pm 0.010$ & $0.13 \pm 0.03$ & $7.5 \pm 0.7$ & $9.4 \pm 1.2$ & $-0.24 \pm 0.17$  & $1.81 \pm 0.03$ \\
1103010220 & 261.253 & 15 & 104.5 & 0.48 & $1.356 \pm 0.011$ & $0.14 \pm 0.05$ & $8.3 \pm 0.7$ & $10.6 \pm 1.2$ & $-0.02 \pm 0.15$  & $1.76 \pm 0.03$ \\
1103010221 & 262.089 & 14 & 100.5 & 0.47 & $1.563 \pm 0.013$ & $0.21 \pm 0.04$ & $10.1 \pm 0.7$ & $14.2 \pm 1.2$ & $-0.09 \pm 0.12$  & $1.85 \pm 0.03$ \\
1103010223 & 339.690 & 37 & 111.0 & 0.44 & $1.973 \pm 0.008$ & $0.15 \pm 0.02$ & $7.3 \pm 0.5$ & $10.4 \pm 0.8$ & $-0.06 \pm 0.11$  & $1.91 \pm 0.02$ \\
2596010101 & 348.694 & 55 & 142.9 & 0.43 & $2.080 \pm 0.009$ & $0.21 \pm 0.03$ & $6.8 \pm 0.3$ & $11.4 \pm 0.7$ & $-0.13 \pm 0.09$  & $1.96 \pm 0.02$ \\
2596010201 & 355.707 & 54 & 143.4 & 0.45 & $2.018 \pm 0.005$ & $0.15 \pm 0.02$ & $7.9 \pm 0.3$ & $9.4 \pm 0.6$ & $\phantom{-}0.02 \pm 0.08$  & $1.97 \pm 0.02$ \\
2596010301 & 362.662 & 48 & 159.2 & 0.43 & $2.230 \pm 0.007$ & $0.19 \pm 0.02$ & $6.9 \pm 0.3$ & $11.5 \pm 0.6$ & $-0.20 \pm 0.08$  & $2.01 \pm 0.02$ \\
2596010401 & 369.427 & 49 & 163.0 & 0.42 & $2.429 \pm 0.011$ & $0.26 \pm 0.03$ & $7.5 \pm 0.3$ & $10.1 \pm 0.7$ & $-0.19 \pm 0.09$  & $2.02 \pm 0.02$ \\
2596010601 & 383.037 & 27 & 168.1 & 0.44 & $2.205 \pm 0.006$ & $0.17 \pm 0.02$ & $9.2 \pm 0.4$ & $12.4 \pm 0.7$ & $-0.01 \pm 0.08$  & $2.00 \pm 0.02$ \\
2596010701 & 389.157 & 45 & 153.8 & 0.44 & $2.493 \pm 0.017$ & $0.42 \pm 0.04$ & $8.8 \pm 0.3$ & $12.2 \pm 0.7$ & $-0.13 \pm 0.07$  & $2.07 \pm 0.02$ \\
2596010801 & 396.554 & 76 & 120.6 & 0.51 & $2.096 \pm 0.005$ & $0.22 \pm 0.02$ & $10.5 \pm 0.3$ & $12.5 \pm 0.5$ & $-0.06 \pm 0.05$  & $1.95 \pm 0.02$ \\
\noalign{\smallskip}
    \noalign{\smallskip}
  \hline

 \end{tabular}}
 \flushleft{$^a$ Count rate in the $0.3-12.0$~keV band.\\
 $^b$ Ratio between the count rates in the $5-12$~keV and $2-5$~keV energy bands. \\ $^c$ QPO phase-lags between the $5-12$~keV and $2-5$~keV energy bands, using the latter as reference band.}
  \label{tab:tabla}
\end{table*}

\FloatBarrier 
\clearpage

\end{document}